\documentclass{aa}
\usepackage{natbib,graphicx}
\usepackage{amsmath,amsbsy}
\usepackage{txfonts}
\newcommand{\changed}[1]{#1}
\newcommand{\change}[1]{#1}
\begin{document}
   \title{Dust coagulation in protoplanetary disks: porosity matters}

   \author{C.W. Ormel
          \inst{1}
          \and
          M. Spaans
          \inst{1}
          \and
          A.G.G.M. Tielens
          \inst{1,2}
          }

   \offprints{C.W. Ormel}

   \institute{Kapteyn Astronomical Institute, University of Groningen, PO box 800, 9700 AV  Groningen, The Netherlands\\
             \email{ormel@astro.rug.nl}
             \email{spaans@astro.rug.nl}
             \and
             NASA Ames Research Center, Mail Stop 245-3, Moffett Field, CA 94035, USA\\
             \email{tielens@astro.rug.nl}
             }

   \date{Accepted September 27, 2006}

  \abstract
   {\\ \textbf{Context.} Sticking of colliding dust particles through van der Waals forces is the first stage in the grain growth process in protoplanetary disks, eventually leading to the formation of comets, asteroids and planets. A key aspect of the collisional evolution is the coupling between dust and gas motions, which depends on the internal structure (porosity) of aggregates.
    \\ \textbf{Aims.} To quantify the importance of the internal structure on the collisional evolution of particles, and to create a new coagulation model to investigate the difference between porous and compact coagulation in the context of a turbulent protoplanetary disk.
    \\ \textbf{Methods.} We have developed simple prescriptions for the collisional evolution of porosity of grain-aggregates in grain-grain collisions. Three regimes can then be distinguished: `hit-and-stick' at low velocities, with an increase in porosity; compaction at intermediate velocities, with a decrease of porosity; and fragmentation at high velocities. This study has been restricted to physical regimes where fragmentation is unimportant. The temporal evolution has been followed using a Monte Carlo coagulation code. 
    \\ \textbf{Results.} This collision model is applied to the conditions of the (gas dominated) protoplanetary disk, with an $\alpha_\textrm{T}$ parameter characterising the turbulent viscosity. We can discern three different stages in the particle growth process. Initially, growth is driven by Brownian motion and the relatively low velocities involved lead to a rapid increase in porosity of the growing aggregate. The subsequent second stage is characterised by much higher, turbulent driven velocities and the particles compact. As they compact, their mass-to-surface area increases and eventually they enter the third stage, the settling out to the mid-plane. We find that when compared to standard, compact models of coagulation, porous growth delays the onset of settling, because the surface area-to-mass ratio is higher, a consequence of the build-up of porosity during the initial stages. As a result, particles grow orders of magnitudes larger in mass before they rain-out to the mid-plane. Depending on the precise value of $\alpha_\mathrm{T}$ and on the position in the nebula, aggregates can grow to (porous) sizes of $\sim \mathrm{10\ cm}$ in a few thousand years. We also find that collisional energies are higher than in the limited PCA/CCA fractal models, thereby allowing aggregates to restructure. It is concluded that the microphysics of collisions plays a key role in the growth process.

   }
   \maketitle
%

\section{Introduction}
Understanding the formation of planetary systems is one of the central themes of modern astrophysics. New stars form in molecular cloud cores when these cores contract under the influence of gravity. This contraction leads to the formation of a central object surrounded by a disk \citep{1987ARA&A..25...23S}. Planets are generally thought to form in these disks, but neither the precise physical conditions required for, nor the processes involved in planetary body assemblage are well understood. Generally, it is thought that grain growth starts with the sticking of sub-micron-sized grains colliding at low velocities \citep{1993prpl.conf.1031W}. The sticking is then driven by weak van der Waals interaction forces between the grains. Relative velocities may reflect Brownian motion or differences in coupling to turbulence in the disk. Eventually, when the grains have grown to $\sim$cm-sizes, they will settle to the mid-plane of the disk, forming a thin dust layer where further growth to planetesimal sizes can take place \citep{1969Safron,1993prpl.conf.1031W}.

There is much observational support for the growth of dust grains in protoplanetary disks from sub-micron to millimetre size scales. In particular, the contrast of the $10\ \mu\mathrm{m}$ silicate emission feature relative to the local continuum shows that the grain size in the disk's photosphere -- where these features originate -- has increased from sub-micron sizes characteristic for interstellar dust, to the micron-sized range \citep{2003A&A...400L..21V,2003A&A...409L..25M,2003A&A...412L..43P,2006ApJ...639..275K}. Furthermore, observations of the continuum (sub)millimetre emission -- originating from the deeper layers of protoplanetary disks -- show typical grain sizes in the range of millimetres, outside the range of interstellar grain sizes by many orders of magnitudes \citep{1991ApJ...381..250B}. Additional support for the importance of collisional grain growth follows from analytical studies of solar system dust. In particular, many interplanetary dust particles (IDPs) collected at stratospheric altitudes, consist of a large number of small grains assembled in a very open structure as expected for collisional aggregates \citep{1979Metic..14R.358B}. These types of IDPs are thought to derive from comets and, indeed, comets may consist largely of such loosely bound aggregates. In addition, chondrules recovered from meteorites show dust rims which are generally attributed to collisional accretion processes in the solar nebula before the meteorite and its parent body were assembled \citep{1992GeCoA..56.2873M,2004Icar..168..484C}.

The properties of the dust are of key importance to the evolution of protoplanetary disks. First and foremost, planet formation starts at the dust size and the dust characteristics at the smallest sizes will set the table for further growth. Second, the opacity is dominated by absorption and scattering by dust grains. Hence, the radiative transfer, temperature structure, as well as emission spectrum of protoplanetary disks are controlled by the characteristics of the dust \citep{2000prpl.conf..533B,2000A&A...360..213B}. Third, in turn, the temperature structure will dominate the structure of the disk, including such aspects as flaring. Fourth, dust grains provide convenient surfaces that can promote chemical reactions. Specifically, ice mantles formed by accretion and reactions between simple precursor molecules are widespread in regions of star formation \citep{2004ApJS..151...35G,2004ApJS..154..359B}. In fact, grain surfaces may be catalytically active in the warm gas of the inner disks around protostars, converting CO into CH$_4$ \citep{2001M&PS...36...75K}.

Most early studies of the coagulation process and the characteristics of the resulting aggregates assumed hit-and-stick collisions where randomly colliding partners stick at the point of initial contact \citep{1988AnnRevPhysChem...39..237M,1988ApJ...329L..39M,1993A&A...280..617O}. The structure of the aggregate then depends on whether the collision is between a cluster and a monomer (PCA) or between two clusters (CCA). The latter leads to very open and fluffy structures with fractal dimensions less than 2, while the former leads to more compact structures and a fractal dimension (for large aggregates) near 3. \changed{\citet{1993A&A...280..617O} also investigated the pre-fractal limit in which aggregates consist of $\lesssim 1000$ monomers. He provides simple analytical expressions for, e.g., the geometrical and collisional cross-section in the case of PCA and CCA aggregation. These expressions include a structural parameter, $x$, which describes the openness (or fluffiness) of the particle. In this study we also introduce a structural parameter and confirm its importance in coagulation studies. }

Over the last decade much insight has been gained in the structure of collisional aggregates through extensive, elegant, experimental studies \citep{2002AdSpR..29..497B,2004ASPC..309..369B} supported by theoretical analysis \citep{1993ApJ...407..806C,1997ApJ...480..647D}. These studies have revealed the importance of rolling of the constituent monomers for the resulting aggregate structure. At low collision velocities, the hit-and-stick assumption is well justified but at high collision velocities, aggregates will absorb much of the collision energy by restructuring to a more compact configuration. At very high velocities, collisions will lead to disruption, fragmentation, and a decrease in particle mass. The critical velocities separating these collisional regimes are related to material properties such as surface free energy and Young's modulus as well as monomer size and the size of the clusters.

While the porous and open structure of collisional aggregates is well recognised, their importance for the evolution of growing aggregates in a protoplanetary setting is largely ignored. Most theoretical studies represent growing aggregates either by an equivalent sphere \citep[e.g.,][]{1984Icar...60..553W,1988A&A...195..183M,2005ApJ...625..414T,2006ApJ...640.1099N} or adopt the fractal dimension linking mass and size characteristic for CCA or PCA growth \citep[e.g.,][]{1997Icar..127..290W,2001ApJ...551..461S,2005A&A...434..971D}. \citet{1993A&A...280..617O} and \citet{1999Icar..141..388K} explicitly account for aggregates' internal structure in their numerical models, although, due to computational reasons, only a limited growth can be simulated. Indeed, the internal structure of collisional aggregates is the key to their subsequent growth. The coupling of aggregates to the turbulent motion of the gas is controlled by their surface area-to-mass ratio, while the relative velocity between the collision partners dictates in turn the restructuring of the resulting aggregate. Moreover, as a result of the growth process from sub-micron-sized monomers to cm-sized aggregates, the coupling to the gas velocity field may well change due to compaction. Indeed, compaction can be an important catalyst for aggregates to settle out in a mid-plane dust layer. Despite its importance for the collisional growth of aggregates in a protoplanetary environment, the evolution of porosity has not yet been theoretically investigated. The present paper focuses precisely on this aspect of grain growth in protoplanetary disks.

This paper is organised as follows. In Sect.\ \ref{sec:model} a model is presented for the treatment of the porosity as a dynamic variable. This entails defining how porosity\changed{, or rather the openness of aggregates,} is related to the surface area-to-mass ratio (Sect.\ \ref{sec:porosaggl}) as well as quantifying how it is affected by collisions (Sect.\ \ref{sec:colmol}). In Sect.\ \ref{sec:mccode} a Monte Carlo model is presented to compute the collisional evolution, taking full account of the collisional aspect and all features of the porosity model. Section\ \ref{sec:res} then applies the model to the upper regions of the protoplanetary disks. Results from the porous model of this paper are compared to traditional, compact models. In Sect.\ \ref{sec:discuss} we discuss the differences of the new collision model with respect to PCA and CCA models of aggregation and also discuss our results from an observational perspective, before summarising the main results in Sect.\ \ref{sec:concl}.

\section{Collision model\label{sec:model}}
Dust grains are dynamically coupled to the turbulent motions of the gas and `suspended' in the (slowly accreting) protoplanetary nebula. Upon collisions, these small dust grains can stick due to van der Waals forces, forming larger aggregates. Eventually, when these agglomerates have grown very large ($\sim$cm-sized), they will decouple from the gas motions and settle in a thin disk at the mid-plane. Further collisional growth in the mid-plane can then lead to the formation of planetesimals ($\sim$km-sized). Upto this point, growth is driven by weak van der Waals forces, but for km-sized planetesimals gravitational forces take over and rapid growth to a planet takes place.
Here we focus on this process of small grains suspended in the nebula forming larger conglomerates. Coagulation is driven by the relative grain velocities. Velocities and the kinematics of dust in a turbulent nebula are discussed in Sect.\ \ref{sec:environ}. The frictional coupling between dust and gas depends largely on the area-to-mass ratio of the grains and hence on the internal structure of the dust. Section\ \ref{sec:porosaggl} describes the relation between the area-to-mass ratio and the porosity of the dust agglomerates. In Sect.\ \ref{sec:colmol}, we discuss experimental and theoretical studies on the microphysics of dust coagulation and develop a simple model, given in the form of easily applicable recipes, which describes how the mass and porosity of the dust evolve in collisions between two dust agglomerates. In Sect.\ \ref{sec:ccapcacomp}, finally, we compare these recipes in the fractal limit to the frequently used formulations of Particle-Cluster Aggregation (PCA) and Cluster-Cluster Aggregation (CCA).

\subsection{The turbulent protoplanetary disk\label{sec:environ}}
\begin{figure}
  \resizebox{\hsize}{!}{\includegraphics{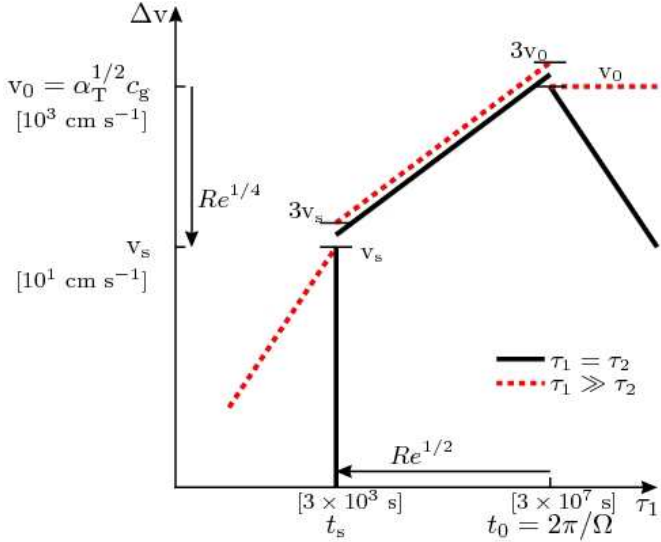}}
  \caption{Sketch of the turbulence induced relative velocities, $\Delta \mathrm{v}$, as function of friction time, $\tau_1$, for equal friction times, $\tau_1 = \tau_2$, (solid line) and different friction times $\tau_2 \ll \tau_2$ (dashed line) according to the analytic fits of Eqs.\ (\ref{eq:delvel1}, \ref{eq:delvel2}) \citep[after][]{1984Icar...60..553W}. $t_0$ and $\mathrm{v}_0$ are set by the global properties of the disk, while the range over which turbulence is important is determined by the Reynolds number, $Re$. Values between brackets denote typical values for $t_0 = 1\ \textrm{yr}$, $c_\mathrm{g} = 10^5\ \mathrm{cm\ s}^{-1}$ and $\alpha_\textrm{T} = 10^{-4}$.}
  \label{fig:deltav}
\end{figure}

For the characterisation of the gas parameters of the protoplanetary disk we use the minimum-mass solar nebula (MSN) model as described by \citet{1981PThPS..70...35H} and \citet{1986Icar...67..375N}. The surface gas density of the disk, $\Sigma_\mathrm{g}$, is assumed to fall off as a $-1.5$ power-law with heliocentric radius ($R$) and the temperature scales as $R^{-1/2}$. The vertical structure of the disk is assumed isothermal, resulting in a density distribution which is Gaussian in the height above the mid-plane, $z$. The scaleheight of the disk, $H_\mathrm{g}$, is an increasing function of radius, $H_\mathrm{g} = c_\mathrm{g}/\Omega \propto R^{5/4}$, with $c_\mathrm{g}$ the local isothermal sound speed and $\Omega$ the Keplerian rotation frequency. The thermal gas motions will induce relative (Brownian) velocities between dust particles with a mean rms-relative velocity of
\begin{equation}
    \Delta \mathrm{v_{Brownian}} (m_1, m_2) = \sqrt{ \frac{8k_\mathrm{B}T(m_1+m_2)}{\pi m_1m_2} },
\end{equation}
with $m_1,m_2$ the masses of the particles and $k_\mathrm{B}$ Boltzmann's constant. Except for low mass \changed{particles of size $\lesssim 10\ \mu\mathrm{m}$}, these velocities are negligibly small \changed{when compared to the} relative velocities \changed{induced by the coupling with the turbulent gas}. We assume that the turbulence is characterised by the \citet{1973A&A....24..337S} $\alpha_\mathrm{T}$-parameter,
\begin{equation}
    \nu_\mathrm{T} = \alpha_\mathrm{T} c_\mathrm{g} H_\mathrm{g} = \alpha_T c_g^2 / \Omega \approx \mathrm{v}_0 \ell_0 = \mathrm{v}_0^2 t_0,
\end{equation}
with $\mathrm{v}_0, \ell_0$ and $t_0$, respectively, the velocity, the size and the turn-over time of the largest eddies. According to the standard (Kolmogorov) turbulence theory, turbulent energy is injected on the largest scales and transported to and eventually dissipated at the smallest eddies -- characterised by turn-over time (or dissipation timescale), $t_s$, velocity, $\mathrm{v}_\mathrm{s}$, and scale size, $\ell_\mathrm{s}$, given by $\nu_\mathrm{m} = \mathrm{v}_s \ell_\mathrm{s}$, with $\nu_\mathrm{m}$ the molecular viscosity.\footnote{$\nu_\mathrm{m} = c_\mathrm{g} \mu m_\mathrm{H}/2\rho_\mathrm{g} \sigma_\mathrm{mol}$ with $\mu$ and $\sigma_\mathrm{mol}$, respectively, the mean molecular weight and mean molecular cross-section of the gas molecules.} We can then define the Reynolds number as, $Re = \nu_\mathrm{T}/\nu_\mathrm{m}$, giving $\mathrm{v}_\mathrm{s} = Re^{-1/4} \mathrm{v}_0 $ and  $t_\mathrm{s} = Re^{-1/2} t_0$.
If $t_0$ is assumed to be (of the order of) the Kepler time \citep{1992A&A...263..387D}, $t_0 = 2\pi/\Omega^{-1}$, the turbulence is fully characterised by the $\alpha_T$-parameter (see Fig.\ \ref{fig:deltav}):
\begin{subequations}
    \begin{equation}
        t_0 = 2\pi \Omega^{-1}, \qquad t_\mathrm{s} = Re^{-1/2} t_0
    \end{equation}
    \begin{equation}
        \mathrm{v}_0 = \alpha_\mathrm{T}^{1/2} c_\mathrm{g}, \qquad  \mathrm{v}_\mathrm{s} = Re^{-1/4} \mathrm{v_0}
    \end{equation}
    \begin{equation}
        \ell_0 = \alpha_\mathrm{T}^{1/2} H_\mathrm{g}, \qquad \ell_\mathrm{s} = Re^{-3/4} \ell_0.
    \end{equation}
\end{subequations}

This specification of turbulence is of importance, for, together with the friction times of two particles, it determines the (average root-mean-square) velocity between the particles, $\Delta \mathrm{v}_{ij}$, which in turn plays a crucial role in both the collision model of this section as well as in the time-evolution model of Sect.\ \ref{sec:colmol}. These relative velocities are a function of the friction time ($\tau_\mathrm{f}$) of the particles -- the time it takes a particle to react to changes in the motion of the surrounding gas -- which in the Epstein limit is\footnote{The Epstein limit holds for particles with sizes smaller than the mean-free-path of the gas, $a < \frac{9}{4}\lambda_\mathrm{mfp}$. If this limit is exceeded, friction times increase by a factor $\frac{4}{9}a/\lambda_\mathrm{mfp}$ and quadratically scale with radius \changed{\citep{1972fpp..conf..211W,1977MNRAS.180...57W,2004ApJ...614..960S}}.}

\begin{equation}
    \tau_\mathrm{f} = \frac{3}{4c_\mathrm{g} \rho_\mathrm{g}} \frac{m}{A},
    \label{eq:frict}
\end{equation}
where $\rho_\mathrm{g}$ is the local mass density of the gas, $m$ the mass of the particle and $A$ its cross-section. In particular, if the friction time of a particle is much smaller than the turnover time of the smallest eddy ($\tau_\mathrm{f} \ll t_\mathrm{s}$), the particle is coupled to \textit{all} eddies and flows with the gas. Therefore, grains with $\tau_\mathrm{f} \ll t_\mathrm{s}$ have highly correlated velocities. Eventually, however, due to growth and compaction, agglomerates will cross the Kolmogorov `barrier' (i.e., $\tau_\mathrm{f} > t_\mathrm{s}$) and are then insensitive to eddies with turnover times smaller than $\tau_\mathrm{f}$, since these eddies will have disappeared before they are capable of `aligning' the particle's motion. At this point, grains can develop large relative motions (Fig.\ \ref{fig:deltav}).

To calculate relative velocities accurately, contributions from all eddies have to be included. \citet{1980A&A....85..316V} started quantifying these effects by dividing eddies into several classes and subsequently integrated over the full Kolmogorov power spectrum. With the exception of some special cases of the particle's friction times \citep{2003Icar..164..127C} in which $\Delta \mathrm{v}$ can be expressed analytically, $\Delta \mathrm{v}$ between particles with different $\tau_\textrm{f}$ can be presented only numerically \citep{1980A&A....85..316V,1991A&A...242..286M}. \citet{1984Icar...60..553W,1984LPI....15..900W}, however, has presented analytical fits, which have been frequently applied in subsequent coagulation models \citep[e.g.,][]{2001ApJ...551..461S,2005A&A...434..971D,2005ApJ...625..414T}. For particle friction times, $\tau_1$ and $\tau_2$ \changed{($\tau_1 \ge \tau_2$)}, less than $t_0$, these read,\footnote{\citet{2005A&A...434..971D} note that the second expression in Eq.\ (\ref{eq:delvel1}) may not exceed $\mathrm{v}_0$.}
\begin{subequations}
\begin{equation}
\Delta \mathrm{v}_{12}\ (\tau_1, \tau_2) = \begin{cases} 
    \mathrm{v}_\mathrm{s} \dfrac{\left( \tau_1 - \tau_2\right)}{t_\mathrm{s}}                & \qquad \tau_1,\tau_2 <t_\mathrm{s} \\[3mm]
    \mathrm{v}_\mathrm{s}  \dfrac{3}{1 + \tau_2/\tau_1}\sqrt{\dfrac{\tau_1}{t_\mathrm{s}}}   & \qquad t_s < \tau_1 <t_0,
    \label{eq:delvel1}
\end{cases} \\     
\end{equation}
where $\tau_1$ is the larger of the two friction times. If $\tau_1$ exceeds $t_0$ the relative velocities become,
\begin{equation}
    \Delta \mathrm{v}_{12} = \begin{cases}
                     \mathrm{v}_0                                           & \qquad \tau_2 < t_0 < \tau_1 \\[3mm]
                     \mathrm{v}_0 \dfrac{(\tau_1+\tau_2)t_0}{2\tau_1\tau_2} & \qquad \tau_1, \tau_2 > t_0.
                    \end{cases}
    \label{eq:delvel2}
\end{equation}
\end{subequations}
For example, in the regime where both friction times are small ($\tau_1, \tau_2 < t_\mathrm{s}$) the turbulence induced relative velocity is negligible when $\tau_2 \approx \tau_1$, but scales linearly with $\tau_1$ when particle 2's mass-to-area ratio is much larger than that of particle 1 (Fig.\ \ref{fig:deltav}). Thus, in the $\tau_1,\tau_2 < t_s$ regime particles with very different $A/m$-ratios will preferentially collide, because differential velocities are then highest. When one of the particles' friction time exceeds $t_\mathrm{s}$ the dependence on the absolute difference in $\tau_\mathrm{f}$ vanishes and the relative velocities now scale with the square root of $m/A$ of the largest $\tau_\mathrm{f}$. As can be seen in Fig.\ \ref{fig:deltav} relative velocities are rather insensitive to the precise ratio of the particles' friction times in this regime. (Because of the simplicity of the expressions for $\Delta \mathrm{v}$ in Eqs.\ (\ref{eq:delvel1}, \ref{eq:delvel2}) the lines do not connect at $\tau_1 = t_s$ and $\tau_1 = t_0$.) In the $\tau_\textrm{f}>t_0$ regime, (Eq.\ \ref{eq:delvel2}) relative velocities would stop increasing and eventually become only a minor perturbation to the motion of the particle. These large $\tau_\textrm{f}$ regimes are, however, not reached in the early stages of coagulation considered in this paper. Here, it is clear that the relative grain velocities are regulated by the area-to-mass ratio of the colliding grains which sets the friction timescale (Eq.\ \ref{eq:frict}) and hence the velocities (Eqs.\ \ref{eq:delvel1}, \ref{eq:delvel2}).
\subsection{Porosity of agglomerates\label{sec:porosaggl}}
\begin{figure*}
  \centering
  \resizebox{\hsize}{!}{
    \includegraphics{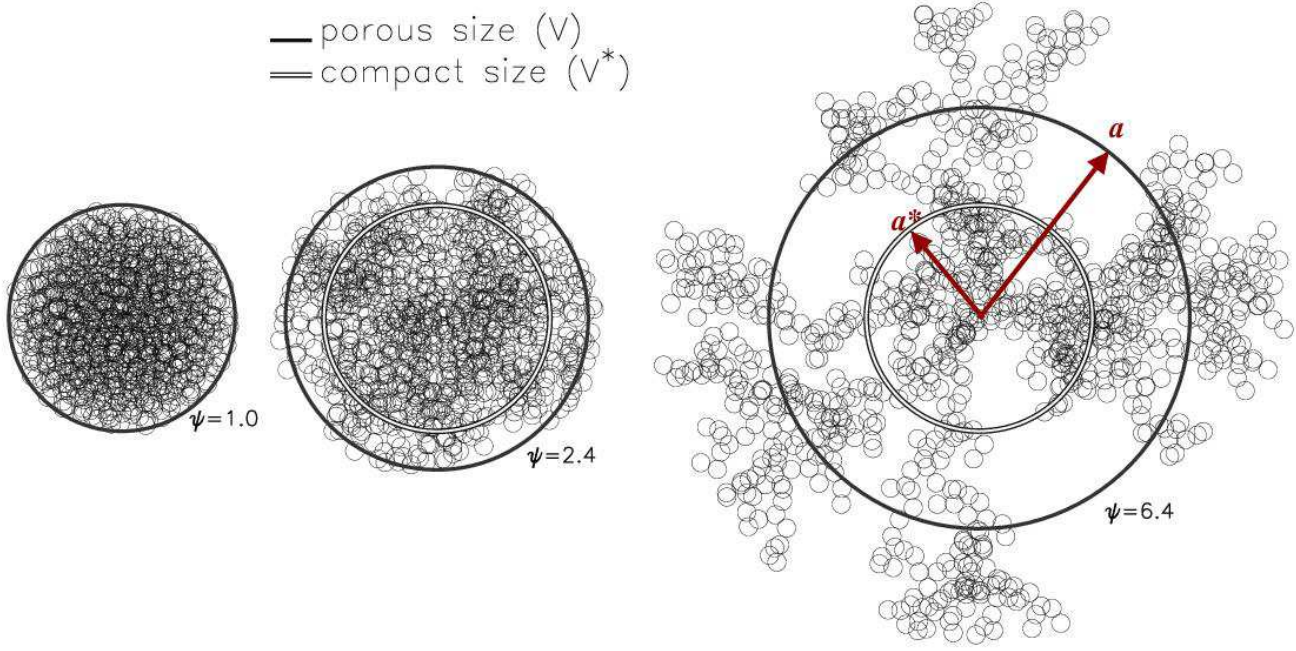}
    }
    \caption{Projections of fractal aggregates illustrating the relation between $A$, $V$ and $V^*$. All fractals consist of 1\,000 monomers and have consequently the same compact volume, $V^*$ (inner circle). The black, outer circle gives the area, $A$, equal to the total projected surface area of the agglomerate. This area subsequently defines the volume, $V$ (e.g., $V \sim A^{3/2}$) and the \changed{enlargement factor} $\psi$ of the aggregate (Eq.\ \ref{eq:psidef}). (\textit{left}) Compact aggregate, defining the compact volume $V^*$. (\textit{centre}) Porous aggregate. The geometric cross-section, $A$ (outer circle), is larger than its compact equivalent (inner circle). (\textit{right}) Even more porous aggregate. Arrows denote the compact and porous radii, $a^*$ and $a$.}
  \label{fig:fractalfig}
\end{figure*}

For compact, solid particles,\footnote{The reader should note that the words `particle', `agglomerate' and `aggregate' are frequently interchanged throughout this and other paragraphs.} the area-to-mass ratio scales linearly with the size. However, for porous aggregates the $A/m$ ratio depends on the internal structure of the aggregates. In this section, we discuss how the internal structure of the aggregates affects collisions and, conversely, how collisions affect the aggregates' internal structure. The internal structure is modelled using the \changed{enlargement} parameter. Here, we \changed{define an \textit{enlargement parameter} $\psi$ that is the ratio between its extended volume, $V$, and its compact volume, $V^*$, i.e.,}
\begin{equation}
    \psi = \frac{V}{V^\ast}, \qquad \psi \ge 1.
    \label{eq:psidef}
\end{equation}
$V^*$ is the volume the material occupies in its compacted state, i.e., when particles are packed most efficiently, and $V$ the total (extended) volume of the particle. While $V^*$ reflects the mass of the particle, $V$ is related to the spatial extent of the aggregate. Here, $V$ has been defined as the volume corresponding to the \changed{geometric} cross-section of the aggregate (see Fig.\ \ref{fig:fractalfig}). Figure\ \ref{fig:fractalfig} shows three aggregates of 1\,000 monomers, such that $V^*$ is the same, though with different degrees of compaction. The (from left to right) decreasing compaction translates to an increased \changed{geometric} cross-section, $A$ (outer circle), of the aggregates and hence to an increased $\psi$. Using the linearity between $V^*$ and $m$ and Eq.\ (\ref{eq:psidef}), two parameters, e.g., $m$ and $\psi$, determine $A$ and, consequently, also determine the coupling with the gas (Eq.\ \ref{eq:frict}).

$\psi=1$ \changed{then corresponds to the enlargement factor} of a compact particle with specific density $\rho_\textrm{s}=m/V^*$. This does not necessarily correspond to a homogeneous particle (of zero porosity); it corresponds, however, to the lowest porosity that can be achieved by re-arranging the constituent particles (monomers) within the aggregate, without destroying this substructure through, e.g., melting. 
\changed{Since $\psi \ge 1$ we refer to $\psi$ as the enlargement parameter or enlargement factor. $\psi$ is related to the structural parameter, $x$, of \citet{1993A&A...280..617O}, as $x \propto \psi^2$. Note, the close relationship between $\psi$ and porosity; a higher $\psi$ means more open aggregates, i.e., higher porosity. Therefore we will also use this more familiar reference for $\psi$, but only in a qualitative sense.}

\changed{Since our enlargement parameter, $\psi$, is defined in order to match the geometrical cross-section, $A$, of a particle, we refer to the resulting radii, $a(\psi)$, as geometrical. The other cross-section of importance in coagulation models is $\sigma$, the collisional cross-section. \citet{1993A&A...280..617O} has consequentially defined a `toothing radius', $a_\textrm{tooth}$, such that $\sigma = \pi (a_\textrm{tooth, 1} + a_\textrm{tooth, 2})^2$ and provides expressions for $a_\textrm{tooth}$ for PCA and CCA aggregates. $\sigma$ is larger than $A$ by a factor of order unity (see also \citet{2004PhRvL..93b1103K}) and also depends slightly on the structure of the aggregates, i.e., whether PCA or CCA. In this paper, however, we are mainly concerned with obtaining a model for $\psi$ and have therefore simply used $a$ for the calculation of the collisional cross-section. Likewise, we have also ignored augmentations of $\sigma$ due to effects such as, e.g., charges and grain asymmetries (see again \citet{1993A&A...280..617O}) and rotation of grains \citep{2006Icar..182..274P}.}

The \changed{aggregates' internal structure} has important consequences for the coagulation rate. The geometric cross-section scales, for example, as $\psi^{2/3}$. More subtle are the effects of \changed{$\psi$} on relative velocities, which, as seen above, depend critically on the $A/m$-ratio of both aggregates. In coagulation models where grains are represented as compact bodies $\tau_\mathrm{f}$ increases linearly with size ($A\propto m^{2/3}; \tau_\mathrm{f} \propto m^{1/3}$). However, in the initial stages of coagulation aggregates stick where they meet -- a process characterised by a build-up of porous, fluffy structures, which can be described by fractals. \citet{1988ApJ...329L..39M} have computed the $A/m$ ratio of an initially monodisperse population and found a profound flattening of this ratio as compared to compact models in which it decreases as $m^{-1/3}$. Often, in the `hit-and-stick' regime \changed{the growth shows fractal behaviour and} the cross-section can be directly parameterised as a power-law, i.e.,
\begin{equation}
    A \propto m^\delta, \qquad \tfrac{2}{3} \le \delta \le 1.
\end{equation}
\changed{The lower limit $\delta=\frac{2}{3}$ corresponds to the growth of compact particles, whereas the upper limit of $\delta=1$ denotes the aggregation of chains or planar structures.} In the cluster-cluster coagulation (CCA) model of \citet{1993A&A...280..617O} $\delta=\delta_\mathrm{CCA}=0.95$, a result that agrees well with the findings of \citet{1999Icar..141..388K}, using N-particle simulations in the Brownian motion regime. More recent models, which also include rotation of aggregates \citep{2006Icar..182..274P} also confirm this exponent. The scaling of friction time \changed{and enlargement factor} with mass then become
\begin{equation}
    \tau_\mathrm{f} \propto m^{1-\delta}, \qquad \psi \propto m^{\frac{3}{2}\delta -1},
    \label{eq:frac}
\end{equation}
such that for compact aggregation models ($\delta=2/3$) $\psi$ stays constant. On the other hand, in models of cluster-cluster aggregation ($\delta \approx 1)$, area scales roughly as mass and $\tau_\mathrm{f}$ stays constant or is only weakly dependent on mass, while $\psi$ increases monotonically. Thus, in CCA relative velocities are rather insensitive to growth. As a result, collisions are also less energetic in CCA models, e.g., as compared to compact aggregation models.

The key to the coagulation process in protoplanetary disks is the coupling of the dust to the turbulent motions of the gas and the resulting velocity distribution. In essence, the \changed{enlargement} parameter $\psi$ provides a relationship between mass and surface area for growing aggregates which controls this gas-dust coupling. Equation\ (\ref{eq:frac}) provides a relation for the evolution of $\psi$, but this relation is only valid in certain specific aggregation models, e.g., CCA-coagulation, where similar, equally sized aggregates meet. In reality, however, collisions between particles of all kinds of sizes will occur, although, dependent on the parameters that determine $\Delta \mathrm{v}$, some are just more probable than others. In the end, the growth of grains in protoplanetary disks is controlled by the individual collisions between two aggregates. Therefore, we have to provide prescriptions for the outcome of all possible collisional encounters, i.e., all relevant combinations of $m, \psi$ and $\Delta \mathrm{v}$.

\subsection{The collision model\label{sec:colmol}}
We consider a collision between two particles with the aim of applying the results to a true coagulation model. Essentially, we have to provide a recipe for the \changed{enlargement factor}, $\psi$, after the collision. Two relevant parameter sets enter into the new $\psi$: the progenitor masses and \changed{enlargement factors} (i.e., the $(m_i, \psi_i)$ of the colliding particles). Moreover, the collision energy,
\begin{equation}
  E = \frac{1}{2}\frac{m_1m_2}{m_1+m_2} \Delta \mathrm{v}^2 = \tfrac{1}{2} \mu \Delta \mathrm{v}^2, 
  \label{eq:collen}
\end{equation}
with $\mu$ the reduced mass, is of key importance. At very low velocities, collisions between two aggregates will lead to sticking without internal restructuring, i.e., the particles will stick where they make first contact. At moderate velocities, the internal structure of the resulting aggregate will adjust -- dissipating a major fraction of the kinetic collision energy -- resulting in a compaction of the aggregates. Finally, at very high collision velocities, the colliding aggregates will fragment into smaller units and this can lead to substantial erosion. Following the numerical model of \citet{1997ApJ...480..647D} and the experimental studies by \citet{2000Icar..143..138B}, these collisional regimes are \changed{distinguished by the following critical (collision) energies:}
\begin{itemize}
  \item $E_\mathrm{restr} = 5 E_\mathrm{roll}$, the energy needed for the \changed{onset of compaction};
  \item $E_\mathrm{max-c} \simeq N_\textrm{c} \cdot E_\textrm{roll}$, the energy at which aggregates are maximum compressed. Here, $N_\mathrm{c}$ is the total number of contact surfaces (between monomers) of the agglomerate. For a very open, fluffy agglomerate, $N_\mathrm{c} = N$. With increasing compaction the number of contacts will increase and for a cubic packing $N_\mathrm{c}=3N$. Here, for simplicity, we will adopt $N_\mathrm{c} = N$.
  \item $E_\mathrm{frag} \simeq 3 N_\mathrm{c} \cdot E_\mathrm{break}$, the energy needed for (the onset of) fragmentation of the agglomerate. \changed{Here $E_\mathrm{break}$ is the energy required to break a bond between two monomers. Its magnitude is of similar order as $E_\mathrm{roll}$.}
\end{itemize}
For monomers of the same size $E_\mathrm{roll}$ is given by \citep{1997ApJ...480..647D,2000Icar..143..138B}
\begin{equation}
    E_\mathrm{roll} = 3\pi^2 \gamma a_0 \xi_\mathrm{crit} = \tfrac{1}{2} \pi a_0 F_\mathrm{roll},
\end{equation}
with $a_0$ the radius of the monomer, $\gamma$ the specific surface adhesion energy and $\xi_\mathrm{crit}$ the yield displacement at which rolling commences. $F_\mathrm{roll}$, the rolling force, was measured by \citet{1999PhRvL..83.3328H} to be $F_\mathrm{roll} = (8.5\pm1.6)\times10^{-5}\ \mathrm{dyn}$ for uncoated $\mathrm{SiO}_2$-spheres of surface energy density $\gamma = 14\pm2\ \mathrm{ergs\ cm}^{-2}$. We adopt this value for $F_\mathrm{roll}$ and assume proportionality with $\gamma$ when applying it to other materials. The monomer size, $a_0$, is also an important model parameter directly affecting the outcome of the collision at a given mass; a smaller $a_0$ provides more contacts (for the same mass) and the agglomerate is more resistant to breakup. With these energy thresholds, three qualitatively different collision regimes can then be discerned (see Fig.\ \ref{fig:colenerg}):
\begin{figure}
  \resizebox{\hsize}{!}{\includegraphics{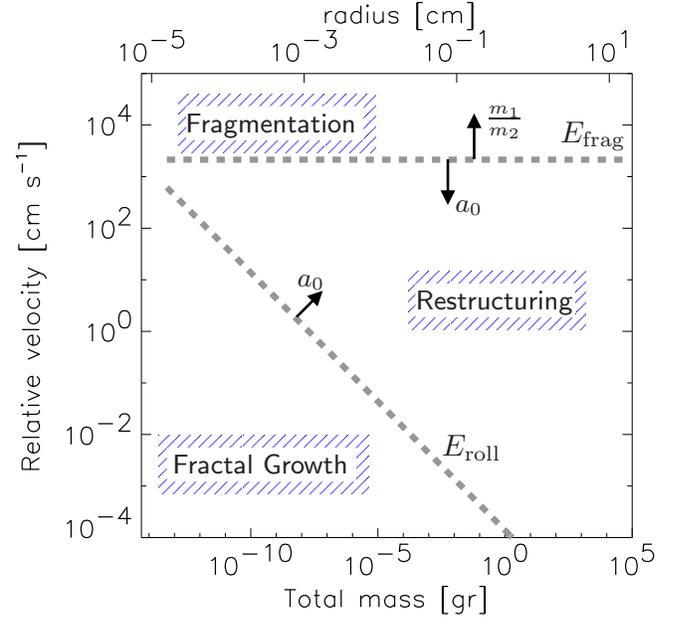}}
  \caption{The collision regimes as function of total particle mass and relative velocity. Thick dashed lines indicate the critical energies for the onset of rolling and fragmentation. Parameters are that of quartz particles ($\rho_\mathrm{s}=3.0\ \mathrm{g\ cm^{-3}}, \gamma=25.0\ \mathrm{ergs\ cm^{-2}}, a_0=0.1\ \mu\mathrm{m}$) and  for the $E_\mathrm{frag}$ line we have assumed equal masses ($m_1=m_2$) \changed{and $E_\textrm{break} = E_\textrm{roll}$}. Arrows indicate how the critical energy lines shift with increasing monomer size and mass-ratio, $m_1/m_2$.}
  \label{fig:colenerg}
\end{figure}
\begin{itemize}
\item[\textit{i}] $E < E_\mathrm{restr}$: No internal restructuring. The particles stick where they meet (`hit-and-stick' growth) as in traditional, lattice-based, aggregation models \citep[e.g.,][]{1988AnnRevPhysChem...39..237M} -- a process leading to fractal aggregates.
\item[\textit{ii}] $E_\mathrm{restr} < E < E_\mathrm{frag}$: Restructuring (compaction) of aggregates. 
\item[\textit{iii}] $E > E_\mathrm{frag}$. Initially, loss of monomers, but for high energies complete fragmentation (e.g., catastrophic collision). This phase requires a recipe for the mass distribution of the fragments. In this paper, fragmenting collisions are avoided by, e.g., `choosing' moderate $\alpha_\textrm{T}$ and stopping coagulation for particles that reach a critical friction time (see Sect.\ \ref{sec:res}). Within these constraints, our results are therefore not compromised by ignoring the fragmentation issue.
\end{itemize}
From Fig.\ \ref{fig:colenerg} it is clear that, when starting with small particles, growth will initially be in the fractal regime. This fractal growth will be followed by a period in which the collisions will promote the restructuring of the growing agglomerate. Fragmentation becomes only important for velocities in excess of $10^3\ \mathrm{cm\ s^{-1}}$. Since we assume that every contact can absorb a unit energy $E_\mathrm{roll}$, the $E_\mathrm{frag}$ line in Fig.\ \ref{fig:colenerg} is independent of total mass. The $\Delta \mathrm{v} \simeq 10^3\ \mathrm{cm\ s^{-1}}$ limit translates to a critical value for the turbulent $\alpha_\textrm{T}$ parameter: $\alpha_T \lesssim 10^{-2}$. Within Fig.\ \ref{fig:colenerg}, the precise `growth path' of the agglomerate will be controlled by evolution of the relative velocities and hence $A/m$ or, equivalently, $\psi$. We will now construct recipes for $\psi$ in, respectively, $i$) the fractal and $ii$) the compaction regime.
\subsubsection{$E < E_\textrm{restr}$: hit-and-stick\label{sec:hitandstick}}
\changed{Besides the usual CCA and PCA formalisms, there have been a few attempts to give prescriptions for the evolving internal structure of aggregates in the hit-and-stick regime. \citet{2001Aerosol..32..1399K} discuss a formalism for obtaining the new fractal dimension in terms of the sizes and fractal dimensions of the two colliding aggregates. One point is, however, that, apart from the fractal dimension, another parameter -- the prefactor -- is needed to fully describe the fractal, although it is usually of order unity. \citet{1993A&A...280..617O}, like this study, introduces only one structural parameter and interpolates between the CCA and PCA limits. We will also follow this idea, but use a different interpolation mechanism. }

\changed{We recognise that in the pure sticking regime} most collisions are between evolved, fluffy aggregates, since the size distribution evolves progressively toward larger sizes. For low velocity-mass combinations (Fig. \ref{fig:colenerg}), where restructuring is unimportant, the collisional growth then resembles the CCA growth process the most. We therefore simply rewrite the fractal law in terms of the individual masses of the particles and keep the CCA exponent,
\begin{equation}
  A = A_1 \left( \frac{m_1+m_2}{m_1} \right)^{\delta_\textrm{CCA}}, \qquad m_1 > m_2.
  \label{eq:fracgrowth1}
\end{equation}
Although collisions between different particles are included in Eq.\ (\ref{eq:fracgrowth1}), we still adopt the CCA-characteristic exponent ($\delta_\textrm{CCA} = 0.95$) to ensure that for the `pure' CCA case ($m_1=m_2;\ A_1=A_2$) this prescription is in accordance with detailed numerical studies \citep{1988ApJ...329L..39M,1993A&A...280..617O,1999Icar..141..388K,2006Icar..182..274P}. There is, however, a modification to Eq.\ (\ref{eq:fracgrowth1}) that must be made. The term in brackets in Eq.\ (\ref{eq:fracgrowth1}) determines the amount of increase in $A$ in the fractal regime. Because fractal growth results from inefficient packing of voluminous objects, it is clear that this term must include parameters describing the spatial extent of the collision partners. These cannot, however, be given by the masses of the particles, since $m$ (alone) does not reflect a spatial size. For example, if we would replace one of the aggregates by one of the same mass, but of lower porosity (i.e., a more compact aggregate), its volume is obviously smaller and packing becomes more efficient. These effects, however, are not reflected in Eq.\ (\ref{eq:fracgrowth1}). For these reasons, we replace $m$ by $V$ in Eq.\ (\ref{eq:fracgrowth1}),  

\begin{equation}
  A = A_1 \left( \frac{V_1+V_2}{V_1} \right)^{\delta_\textrm{CCA}}, \qquad V_1 > V_2.
  \label{eq:fracgrowth2}
\end{equation}
Note that for particles of the same internal density (porosity) Eq.\ (\ref{eq:fracgrowth2}) and Eq.\ (\ref{eq:fracgrowth1}) agree, such that Eq.\ (\ref{eq:fracgrowth2}) also can be seen as an extrapolation from the CCA case, but one that takes account of the different internal structures of the collision partners. Using the relation $A\propto V^{2/3}$ and Eq.\ (\ref{eq:psidef}), Eq.\ (\ref{eq:fracgrowth2}) can be re-written in terms of $m$ and $\psi$ only 
\begin{equation}
  \psi = \langle \psi \rangle_m \left( 1 + \frac{m_2\psi_2}{m_1\psi_1} \right)^{\frac{3}{2}\delta_\textrm{CCA}-1},
    \label{eq:newpor2}
\end{equation}
with $\langle \psi \rangle_m$ the mass-averaged \changed{enlargement factor} of the collision partners,
\begin{equation}
  \langle \psi \rangle_m \equiv \frac{m_1 \psi_1 + m_2 \psi_2}{m_1 + m_2}.
  \label{eq:massaverpor}
\end{equation}
In CCA coagulation ($m_1=m_2$ and $\psi_1 = \psi_2$) we recover Eq.\ (\ref{eq:frac}), but Eq.\ (\ref{eq:newpor2}) now includes all collisions in the hit-and-stick regime. For example, if a large, fluffy aggregate sticks to a compact one, the \changed{enlargement factor} of the newly formed aggregate is higher than the mass-averaged \changed{enlargement factor} of the progenitor particles, $\langle \psi \rangle_m$, but smaller than that of the fluffy collision partner. In Sect.\ \ref{sec:ccapcacomp}, it will be shown, however, that the $\langle \psi \rangle_m$ term underestimates the porous growth when one of the particles is very small, i.e., in PCA-like collisions. This is solved by adding to Eq.\ (\ref{eq:newpor2}) a term that compensates for these cases and our final recipe then becomes
\begin{equation}
  \psi = \langle \psi \rangle_m \left( 1 + \frac{m_2\psi_2}{m_1\psi_1} \right)^{\frac{3}{2}\delta_\textrm{CCA}-1} + \psi_\mathrm{add},
  \label{eq:porfinal}
\end{equation}
where $\psi_\textrm{add}$, a term important only for small $m_2$, is explained in Sect.\ \ref{sec:ccapcacomp}.

\subsubsection{$E>E_\textrm{restr.}$: compaction\label{sec:compcol}}
\begin{figure}
  \resizebox{\hsize}{!}{\includegraphics{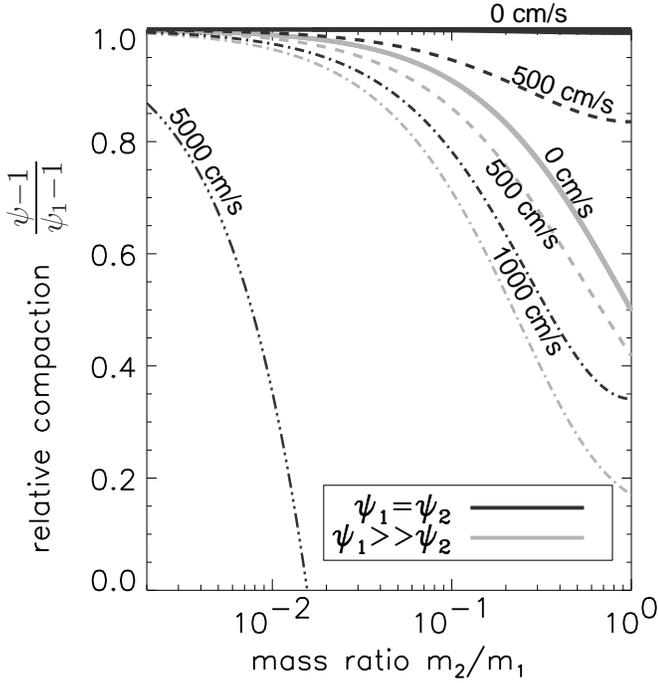}}
  \caption{Relative compaction as function of the mass ratio, $m_2/m_1$, at various collisional velocities. \change{The expression on the y-axis measures the compaction relative to particle 1 and is a function of mass ratio, $m_2/m_1$, only (see Eq.\ \ref{eq:newpor})}. Here $\psi_1$ is the \changed{enlargement factor} of the most massive aggregate, i.e., $m_1 > m_2$, and plots are shown for $\psi_2 \ll \psi_1$ (grey lines) and $\psi_2 = \psi_1$ (black lines). At low mass ratios the curves converge.}
  \label{fig:comp}
\end{figure}

In the compaction limit monomers are restructured at the expense of the porous volume of the aggregates. Following the theoretical study of \citet{1997ApJ...480..647D} we will assume that the (relative) amount of compaction, $\Delta V_\mathrm{p}/V_\mathrm{p}$, scales linearly with collisional energy, i.e., $\Delta V_\mathrm{p}/V_\mathrm{p} = -f_C = E/E_\textrm{max-c} = -E/(N \cdot E_\mathrm{roll})$, where $V_p = V - V^*$ denotes the porous volume within the aggregate and $N$ is the total number of monomers present.\footnote{\changed{Note that we start compaction already at $E=E_\textrm{roll}$ instead of $E=5E_\textrm{roll}$. We have found, however, that the simulations are insensitive to the precise energy at which compaction starts.}} Essentially, this implies that the collision energy is used to set individual monomers in an agglomerate rolling and that this rolling is only stopped when an additional contact is made, resulting in compaction. Recalling that $V = \psi V^*$ and $V_\mathrm{p} = V - V^* = (\psi - 1) V^*$ with $V^*$ proportional to mass, we then find that the porous volume after colliding is
\begin{equation}
    (V_1^* + V_2^*) (\psi - 1) = (1 - f_C) \left[ V_1^* (\psi_1 -1) + V_2^* (\psi_2 -1) \right].
\end{equation}
And the new \changed{enlargement factor}
\begin{align}
    \psi-1 &= (1-f_C) \cdot \frac{1}{m_1+m_2} \left( m_1 (\psi_1-1) + m_2 (\psi_2-1) \right) \nonumber \\ 
             &=(1 - f_C) \left( \langle \psi \rangle_m -1 \right),
             \label{eq:newpor}
\end{align}
with $\langle \psi \rangle_m$ again the mass-averaged \changed{enlargement factor} of the (two) collision partners (Eq.\ \ref{eq:massaverpor}). We illustrate Eq.\ (\ref{eq:newpor}) in Fig.\ \ref{fig:comp} for the limiting cases of equal porosity collisions ($\psi_1 = \psi_2$; black lines) and very porous vs. compact particle collisions ($\psi_1 \gg \psi_2$; grey lines). Higher mass-ratios give higher collision energies (higher $\mu$) and hence more compaction. If velocities are low, ($\mathrm{v} < 100\ \mathrm{cm\ s^{-1}}$) the net compaction occurs primarily through the $\langle \psi \rangle_m$ factor and the curves converge on the $0\ \mathrm{cm\ s^{-1}}$ (thick) line. Only when $\mathrm{v} > 100\ \mathrm{cm\ s^{-1}}$ does the $f_C$-factor start to become important. Collisions at velocities higher than $1\,000\ \mathrm{cm\ s^{-1}}$ can result in fragmentation if the mass-ratios are similar. For low mass ratios the $\langle \psi \rangle_m$ is determined by $\psi_1$ (the highest mass) and there is little difference between the two limiting cases.

\subsection{\label{sec:ccapcacomp}Porosity increase in the PCA and CCA limits}
\begin{figure}[t]
  \resizebox{\hsize}{!}{\includegraphics{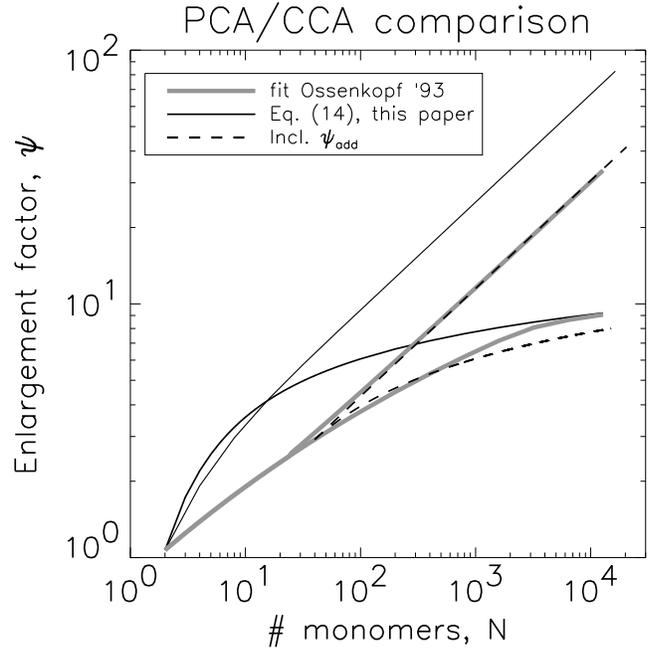}}
  \caption{The \changed{enlargement factor} in the PCA and CCA limits as a function of the number of monomers ($N$) of the particle. The thick grey lines show the fits of \citet{1993A&A...280..617O} for CCA (top) and PCA (bottom) coagulation. The solid line corresponds to the limiting cases of Eq.\ (\ref{eq:newpor2}). The dashed lines show the same limits, but now the zero point lies at $N=40$ (after which the PCA/CCA curves of \citet{1993A&A...280..617O} diverge in porosity) and includes the $\psi_\mathrm{add}$ correction term (Eq.\ \ref{eq:psiadd}).}
  \label{fig:agglfits}
\end{figure}

In the discussion on the $\psi$ recipes in Sect.\ \ref{sec:hitandstick} we have used the CCA fractal exponent ($\delta_\textrm{CCA}=0.95$) as the starting point and extrapolated this empirical relation to Eq.\ (\ref{eq:newpor2}): a general recipe for all collisions in the `hit-and-stick' regime. By definition, CCA-collisions are incorporated into this recipe and we may expect Eq.\ (\ref{eq:newpor2}) also to account for collisions between aggregates having about the same size. The PCA model, where one monomer collides with an agglomerate, is the opposite extreme. If we take the PCA-limit of Eq.\ (\ref{eq:newpor2}), i.e., $m_1 \gg m_2 \sim m_0$, with $m_0$ the monomer mass, the \changed{change in $\psi$} after addition of a monomer becomes
\begin{equation}
  \psi_{N+1} \approx \psi_N + \frac{N_2}{N}\left( \tfrac{3}{2}\delta_\textrm{CCA}\psi_2 - \psi_N \right); \qquad N\gg N_2,
  \label{eq:pcalim}
\end{equation}
with $N=N_1=m/m_0$ the number of monomers of the PCA agglomerate and $N_2=1$ for monomers. Thus, $\psi_N$ goes to $\tfrac{3}{2} \delta_\textrm{CCA} \psi_2 \approx 1.5$ in the PCA asymptotic limit ($N_2 =1; \psi_2=1$).\footnote{Here we take $\psi_2 =1$ as the \changed{enlargement factor} of single monomers.} The fact that an asymptotic limit is reached can be understood intuitively, since there must be a point at which the inward penetration of monomers into the centre of the aggregate, which decreases $\psi$, starts to balance the porous growth due to hit-and-stick collisions at the surface. The asymptotic limit of $\psi \approx 1.5$, however, is much lower than typical PCA models indicate ($\psi \approx 10$) as is illustrated in Fig.\ \ref{fig:agglfits}, where the PCA/CCA limits of our model (solid lines) are compared to detailed numerical simulations of \citet{1993A&A...280..617O} (thick lines). Equation\ (\ref{eq:newpor2}) thus underpredicts the porous growth for PCA-like collisions in which one of the particles is small; a result not too surprising since it originates from the CCA fractal law (Eq.\ \ref{eq:fracgrowth1}), which is constructed to apply only for similar (i.e., equal-sized) particles. For these reasons we add to Eq.\ (\ref{eq:newpor2}) a term that compensates the $-N_2 \psi_N / N$ term in Eq.\ (\ref{eq:pcalim}),
\begin{equation}
  \psi_\textrm{add} = B \frac{m_2}{m_1} \psi_1 \exp \left[ -\mu/ m_\mathrm{F} \right],
  \label{eq:psiadd}
\end{equation}
where the exponential ensures $\psi_\textrm{add}$ is unimportant for collisions between particles well above a certain mass-scale, $m_\mathrm{F}$. With $B=1.0$ and $m_\mathrm{F}=10\ m_0$ we find good correspondence with the results of \citet{1993A&A...280..617O}. In Fig.\ \ref{fig:agglfits} the new CCA and PCA limits are shown by the dashed curves, \changed{where we have shifted the `zero point' from $N = 1$ to $N = 40$, i.e, the starting point for $\psi$ after which we recursively apply Eq.\ (\ref{eq:porfinal}). The transition toward fractal behaviour emerges only after this point and we therefore directly take $\psi$ from the results of \citet{1993A&A...280..617O} when $N \lesssim 40$.}

Although, with Eq.\ (\ref{eq:psiadd}) for $\psi_\mathrm{add}$, Eq.\ (\ref{eq:porfinal}) does achieve the right PCA/CCA fractal limits, we do not claim they actually provide a model for $\psi$. The collision recipes are based on empirical findings and extrapolations from these. However, in contrast to the CCA and PCA limiting collisional growth models, where $\psi$ can be directly parameterised in a single exponent, we recognise that the internal structure of the aggregates is changed during -- and caused by -- the collisional growth process. This is the main qualitative difference of our collision model captured in Eq.\ (\ref{eq:porfinal}). This equation, together with Eq.\ (\ref{eq:newpor}), provides recipes for the collisional evolution of the \changed{enlargement factor}, which can be easily incorporated into time-dependent coagulation models. We acknowledge this is an active area of research and future model efforts may well improve on the present formulation.

\section{Monte Carlo Coagulation\label{sec:mccode}}
\subsection{Outline}
\begin{figure*}
  \resizebox{\hsize}{!}{
    \includegraphics{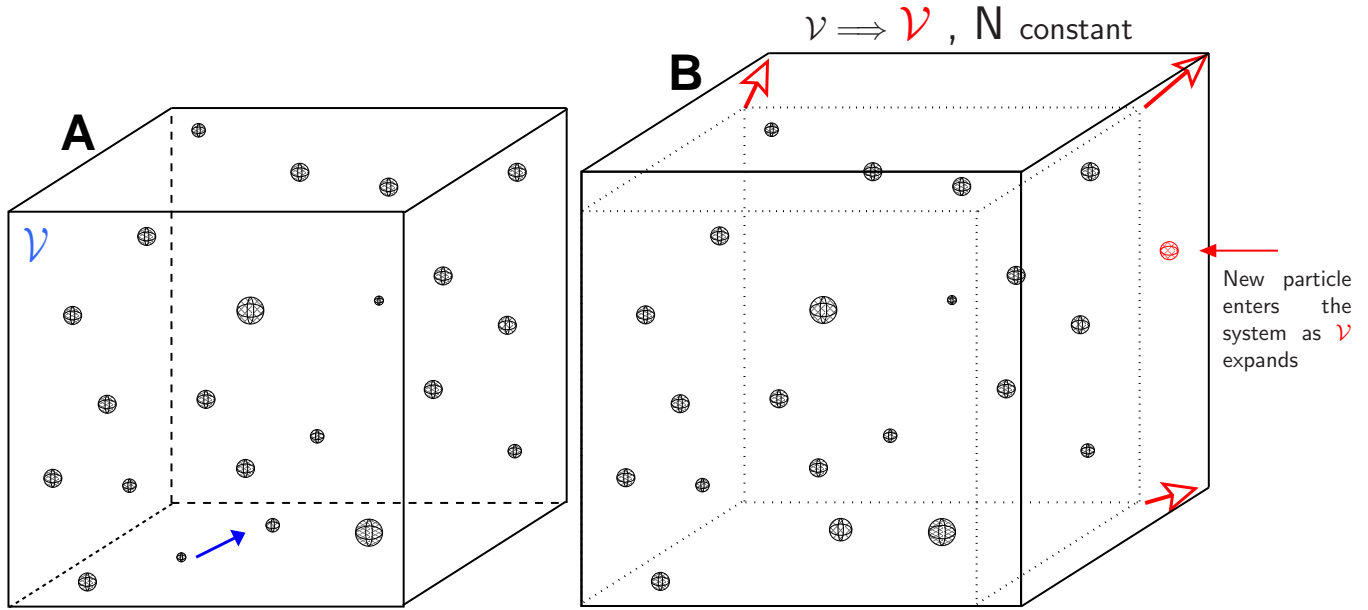}
  }
  \caption{Illustration of the adopted Monte Carlo technique. (A) $N$-particles are assumed to be uniformly distributed in a box of volume ${\cal V}$. The collision rate between particle $i$ and $j$ is $C_{ij} = \sigma_{ij} \Delta \mathrm{v}_{ij} / {\cal V}$. A random number determines which two particles will collide. (B) After the collision  the number of particles is restored by duplicating one of the remaining $N-1$ particles. Physically, this can be interpreted as an expansion of ${\cal V}$ (heavily exaggerated in this figure) to a volume at which ${\cal V}$ contains $N$-particles again. The hypothesis of this method is that the collisional evolution within ${\cal V}$ is representative for the coagulation of the total space under consideration.}
  \label{fig:cubes}
\end{figure*}
To determine the implications the new collision model has for the solar nebula, e.g., as compared to collision models where mass is the only parameter, it must be embedded in a coagulation model that evolves the particle distribution function, $f(\mathbf{x})$. $f(\mathbf{x},t)$ gives the number density of particles with a set of properties (parameters) $\left\{ x_i \right\}$ at time $t$. In compact coagulation models all properties depend only on mass, so $f(\mathbf{x},t)=f(m,t)$. In the model described in Sect.\ \ref{sec:model}, however, the particle's \changed{enlargement factor} has been included as an independent parameter such that $f$ becomes a function of three variables, $f(\mathbf{x},t) = f(m,\psi,t)$. The coagulation equation which describes the evolution of $f(\mathbf{x})$ is
\begin{align}
    \label{eq:smol2d}
              &\hspace*{-5mm}\frac{\partial f(\mathbf{x},t)}{\partial t} =- f(\mathbf{x},t) \int \textrm{d}\mathbf{x'} f(\mathbf{x'},t) K(\mathbf{x,x'}) +\\
    \nonumber &\hspace*{-5mm}+ \frac{1}{2} \int \textrm{d}\mathbf{x'} \textrm{d}\mathbf{x''} f(\mathbf{x}',t) f(\mathbf{x}'',t) K(\mathbf{x',x''},t)\ \delta_\mathrm{k} \left( \mathbf{x} - \Gamma \left(\mathbf{x',x''}) \right) \right),
\end{align}
with $K = \sigma \Delta \mathrm{v}$ the collision kernel, $\delta_\mathrm{k}$ the Kronecker $\delta$-function and $\Gamma$ the collision function, which maps in the case of sticking $2n$ parameters (those of $\mathbf{x'}$ and $\mathbf{x''}$) to $n$, where $n$ is the number of independent parameters with which a particle is characterised. Equation\ (\ref{eq:smol2d}) of course is just an extension of the Smoluchowski equation\footnote{Which reads: 
\begin{align}
    \frac{\partial f(m)}{\partial t} = &- f(m) \int \textrm{d}m'\ K(m,m') f(m') + \nonumber \\
                                       & \frac{1}{2} \int \textrm{d}m'\ K(m',m-m') f(m') f(m-m'),
\end{align}
describing losses of $m$ due to all collisions with $m$ (first term on right hand side) and gains in the distribution of $m$ due to collisions between $m'$ and $m-m'$ (second term), where the factor $\frac{1}{2}$ ensures collisions are not twice accounted for. \citet{1993A&A...280..617O} provides a general extension of the Smoluchowski equation including source and sink terms.} \citep{1916ZPhy...17..557S} to more than one dimension. Applied to the formalism in Sect.\ \ref{sec:model} it describes the loss of particles in state $\mathbf{x} = (m,\psi)$ due to collisions with any other particle (first term) and the gain of `$\mathbf{x}$-particles' that happen to be formed out of any suitable collision between two other particles (second term). Applied to the findings in Sect.\ \ref{sec:model}, $\Gamma$ symbolises the collision recipes with $\Gamma(m_1,\psi_1;m_2,\psi_2) = (m_1+m_2, \psi)$ and $\psi$ is given by Eq.\ (\ref{eq:porfinal}) or Eq.\ (\ref{eq:newpor}), dependent on the collisional energy.

One approach to implement coagulation is to numerically integrate the Smoluchowski equation. However, it is immediately clear that numerically integrating Eq.\ (\ref{eq:smol2d}) becomes a daunting exercise. Integrating the ordinary (1-dimensional) Smoluchowski equation is already a non-trivial matter. Problems of near cancellation (the gain terms often equal the loss terms), mass conservation (systematic propagation of errors) and the problem concerning the determination of a time-step must all be tackled. (\citet{2005A&A...434..971D} in their Appendix B give an overview of the subtleties involved.) 
The $\Gamma$-factor in Eq.\ (\ref{eq:smol2d}) gives a further complication since there is no such thing as `conservation of porosity' and the $\delta$-factor cannot be easily integrated away. \changed{Although there is no fundamental reason against the binning method -- see, e.g., \citet{1993A&A...280..617O} who solves the Smoluchowski equation in two dimensions -- these issues make the whole procedure quite elaborate.} We felt that much of the simplicity of the collision model of Sect.\ \ref{sec:model} would be `buried' by numerical integration of a 2d-Smoluchowski equation and therefore have found it suitable to opt for an approach using direct Monte-Carlo simulation (DSMC) techniques.

The simplicity of using Monte-Carlo methods for coagulation problems is appealing. Basically, $N$-particles are distributed over a volume ${\cal V}$ (see Fig.\ \ref{fig:cubes}A). The evolution then boils down to the determination of the two particles which are involved in the \textit{next} collision. We hereafter assume that the particles are well mixed, i.e., no potential is present, such that the determination of the next collision is governed by basic stochastic principles. 
Then the probability of a collision between particles $i$ and $j$ is given by the collision rate, $C_{ij} = K_{ij} / {\cal V}$ in which $K_{ij}$ is the collision kernel. A random number determines which of the $\frac{1}{2}N(N-1)$ possible collisions will be the next. The collision then creates a new particle, after which the $\{C_{ij}\}$ must be updated and the procedure repeats itself.

The advantages of such an approach are obvious. Most striking perhaps is the `physical character' of Monte-Carlo simulations. The growth-evolution of individual particles is directly traced and can be studied. The algorithm does not use the distribution function, $f$, in a direct way; it is obtained indirectly by binning the particles. Secondly, the above described method is exact, i.e., no `time vs.\ accuracy' considerations have to be made in choosing the time-step $\Delta t$; instead, $\Delta t$ -- the inter-collision time -- is an outcome of the stochastic coagulation process as it is in nature. Furthermore, due to its stochastic nature, no Monte-Carlo simulation is the same. A series of (independent) runs gives at once a measure of the statistical spread in the distribution. Note that the fluctuations around the average are a combination of real stochastic noise and random noise, but it is qualitatively different from the Smoluchowski approach, which describes the evolution of the mean of all possible realisations and is therefore completely deterministic \citep{1977JAtmosSci...32..1977}. From a practical point of view the straightforwardness of the DSMC-method makes that there is no need for resorting to `control parameters' like those required in the numerical-integration method.

The DSMC-method, however, has its limitations. It can be immediately seen, for instance, that when having started with $N$ (say identical) particles of mass $m_0$ in a fixed volume, these will over time pile up in one agglomerate of $m_\mathrm{final}=Nm_0$. The accuracy then steadily decreases during the simulation (in MC-simulations the statistical error scales proportional to $N^{-1/2}$) and most of the computing time is spent in the first few (quite uninteresting) collisions. Consequently, to achieve orders of magnitude growth, the initial number of particles must be extremely large. And because the calculation of the $\{ C_{ij} \}$ goes proportional to $N^2$ (every particle can collide with each other), it becomes clear that this method becomes impracticable. To counter the dependence on large initial particle numbers, one can also try to preserve the number of particles \textit{during} the simulation. This can be done, for instance, by `tossing-up' the particle-distribution when the number of particles reaches $N/2$ as described by \citet{1992JComputPhys...100..116}. A more natural approach perhaps, given by \citet{1998...53..1777}, and adopted here, keeps the number of particles constant at each step; every time a collision takes place one of the remaining $N-1$ particles is randomly duplicated such that the number of particles throughout the simulation stays the same. This procedure can graphically be represented as an increase in the simulated volume, ${\cal V}$ (see Fig.\ \ref{fig:cubes}B), under the assumption that the collisional evolution outside of ${\cal V}$ progresses identically. \citet{1998...53..1777} have shown that the error in the coagulation-scheme now scales logarithmically with the extent of growth, -- or growth factor (GF), defined here as the mean mass over the initial mean mass of the population -- much improved over the constant-${\cal V}$ case, where the error has a square-root dependence on GF. We might worry though about the consequences of the duplication mechanism. It causes a certain degree of `inbreeding', which effects we cannot quantify directly. \citet{1998...53..1777} show it is unimportant for the constant or Brownian kernels used in their studies. However, these kernels are known to behave gently, i.e., they are rather insensitive to irregular changes in the population since the variations in the $K_{ij}$ are small. Perhaps, more erratic kernels are more sensitive to the `duplication mechanism', but we might equally well attribute this sensitivity to the stochastic nature of the coagulation process. At any rate, these consequences are best quantified by running the code multiple times.

\subsection{Implementation\label{sec:mcimplement}}
\begin{figure}
    \includegraphics[scale=0.6]{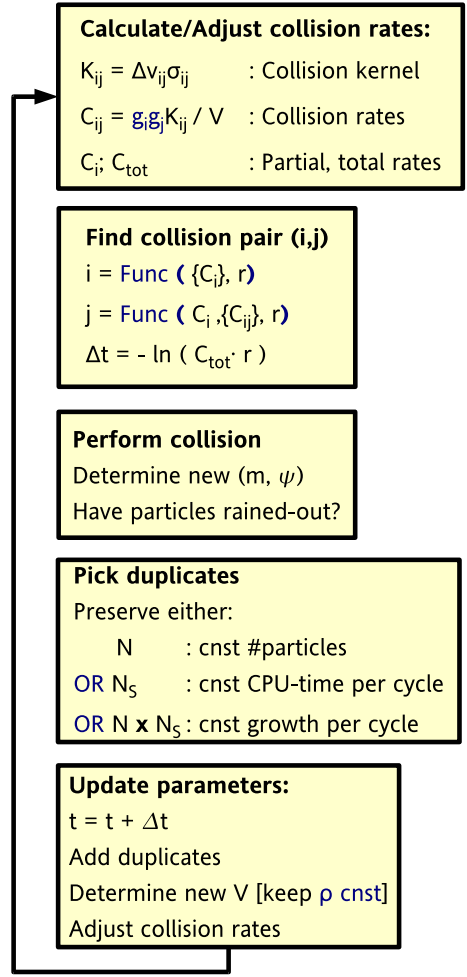}
  \caption{Flow chart of the MC-coagulation method. One cycle corresponds to one collision. `Func' like in $i=\textrm{Func}(\ \{C_{i}\}, r\ )$ indicates that $i$ is a function of the $C_{i}$ values and a random deviate, $r$. The procedure is further explained in the text.}
  \label{fig:flowchart}
\end{figure}

In implementing the DSMC approach we follow the `full conditioning method' of \citet{1977JAtmosSci...32..1977}. This involves the calculation and updating of partial sums, $C_i$,
\begin{equation}
    C_i \equiv \sum_{j=i+1}^{N} C_{ij}, \qquad i=1,\dots,N-1,
    \label{eq:Cisum}
\end{equation}
with $C_{ij} = K_{ij}/{\cal V} = \sigma_{ij}(a_i, a_j) \Delta \mathrm{v}_{ij}(\tau_i, \tau_j) / {\cal V}(\kappa)$ (here $\kappa$ is the total number of collisions since the start of the simulation). These $N-1$ quantities are stored in the memory of the computer. $C_\mathrm{tot} = \sum_i C_i$ is the total coagulation rate. The probability density function, $P(t,i,j)$, i.e., the probability that the \textit{next} collision will occur in time-interval $(t,t+dt)$ $and$ involves particles $i$ and $j$ $(i<j)$ can then be written as \citep{1977JAtmosSci...32..1977}
\begin{align}
    P(t,i,j) &= C_{ij} \exp \left[ -C_\mathrm{tot} t \right] \nonumber \\
             &= \Big( C_\mathrm{tot} \exp \left[ -C_\mathrm{tot} t \right] \Big) \times \Big(  C_i/C_\mathrm{tot} \Big) \times \Big(  C_{ij}/C_i \Big).
\end{align}
Three random deviates, $r_i \in [0,1]$,  then determine successively: \textit{I}) the time it takes until the next collision takes place, $t = C_\mathrm{tot}^{-1} \ln (1/r_1)$; \textit{II}) the first particle ($i$) to collide, by summing over the $C_i$-s (starting with $i=1$) until $r_2 C_\mathrm{tot}$ is exceeded (this fixes $i$); and \textit{III}) its collision partner ($j$) by summing the $C_{ij}$-s over the $j$-index (starting with $j=i+1$) until the value $r_3 C_i$ is exceeded \citep[Eq. 19]{1977JAtmosSci...32..1977}. The outcome of the collision is evaluated using the relevant equations in Sect.\ \ref{sec:model} and the new particle is stored in the $i$-slot. Another random number then determines which of the $N-1$ particles (excluding $j$) will be duplicated and this one is stored in the $j$-slot. Having created, removed and duplicated particles, all of the $C_i$-s need to be updated. This implies only the subtraction/addition of the $C_{ij}$-s that have changed, not the re-computation of Eq.\ (\ref{eq:Cisum}). Moreover, the `duplication procedure' entails a rescaling of the simulated volume, ${\cal V}$, such that the spatial density of solids, $\rho_\mathrm{d} = \sum_i m_i /{\cal V(\kappa)}$ remains constant. The algorithm can then be repeated. All these steps are order-$N$ calculations at worst; most time-consuming are the determination of $j$ (for which $C_{ij}$ has to be calculated) and the update of the $\{ C_i \}$. To achieve a given GF another factor $N$ in computation time is needed,\footnote{Due to the duplication, the mean mass of the system increases with a factor $(N+1)/N$. The growth factor after $\kappa$-steps then becomes
\begin{equation}
    \mathrm{GF} = \left( \frac{N+1}{N} \right)^\kappa.
\end{equation}
Thus, $\ln \mathrm{GF} = \kappa \ln (1+N^{-1}) \approx \kappa/N$ if $N \gg 1$ and $\kappa \approx N \ln \mathrm{GF}$.
} bringing the total CPU-time proportional to $N^2$. These procedures are graphically summarised in Fig.\ \ref{fig:flowchart}.

It is possible, however, to save some CPU time by \changed{taking the duplicates together in the calculation of the collision rates}. If there are $(g_i -1)$ copies of particle $i$, \changed{it would be a waste of time to calculate the (same) rates $g_i$ times. Rather, $g_i$ can be absorbed in the calculation of the (combined) coagulation rate.} $\tilde{C}_{ij} = g_i g_j C_{ij}$ is then the rate at which \changed{one of the $i$-particles} collides with \changed{one of the $j$-particles} ($i\neq j$) and $\tilde{C}_{ii} = \frac{1}{2}g_i(g_i-1) C_{ii}$ \changed{between duplicates}. The CPU time per step is now proportional to $N_\textrm{s}$, the total number of \textit{distinct} particles\changed{, i.e., excluding duplicates}, with $\sum_{i=1}^{N_\textrm{s}} g_i = N$. \changed{To think of it in biological terms: $g_i$ gives the multiplicity of species $i$; $N_\textrm{s}$ the total number of species; and $N$ the total number of living creatures.}

\subsection{Tests}
\begin{figure}
  \resizebox{\hsize}{!}{
    \includegraphics{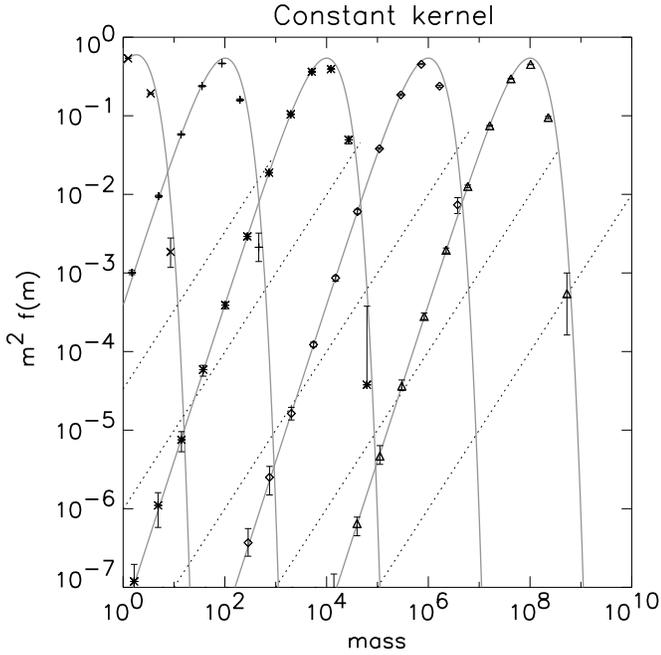}}
    \caption{Test of the Monte Carlo coagulation code -- constant kernel $(K_{ij}=1 )$, 20\,000 particles. At (dimensionless) times $t=1,10^2,10^4,10^6,10^8$ particles are binned and the distribution function $f$ is computed (symbols). The analytical solutions at these times are overplotted by the solid lines. The error bars (hardly visible) show the spread averaged over 10 runs. The dotted lines show the distribution function if all the bins would be occupied by only 1 particle -- it is an auxiliary line with slope 1 showing the lower limit of the (individual) distribution function. }
  \label{fig:test1}
\end{figure}

\begin{figure*}
  \resizebox{\hsize}{!}{
    \includegraphics{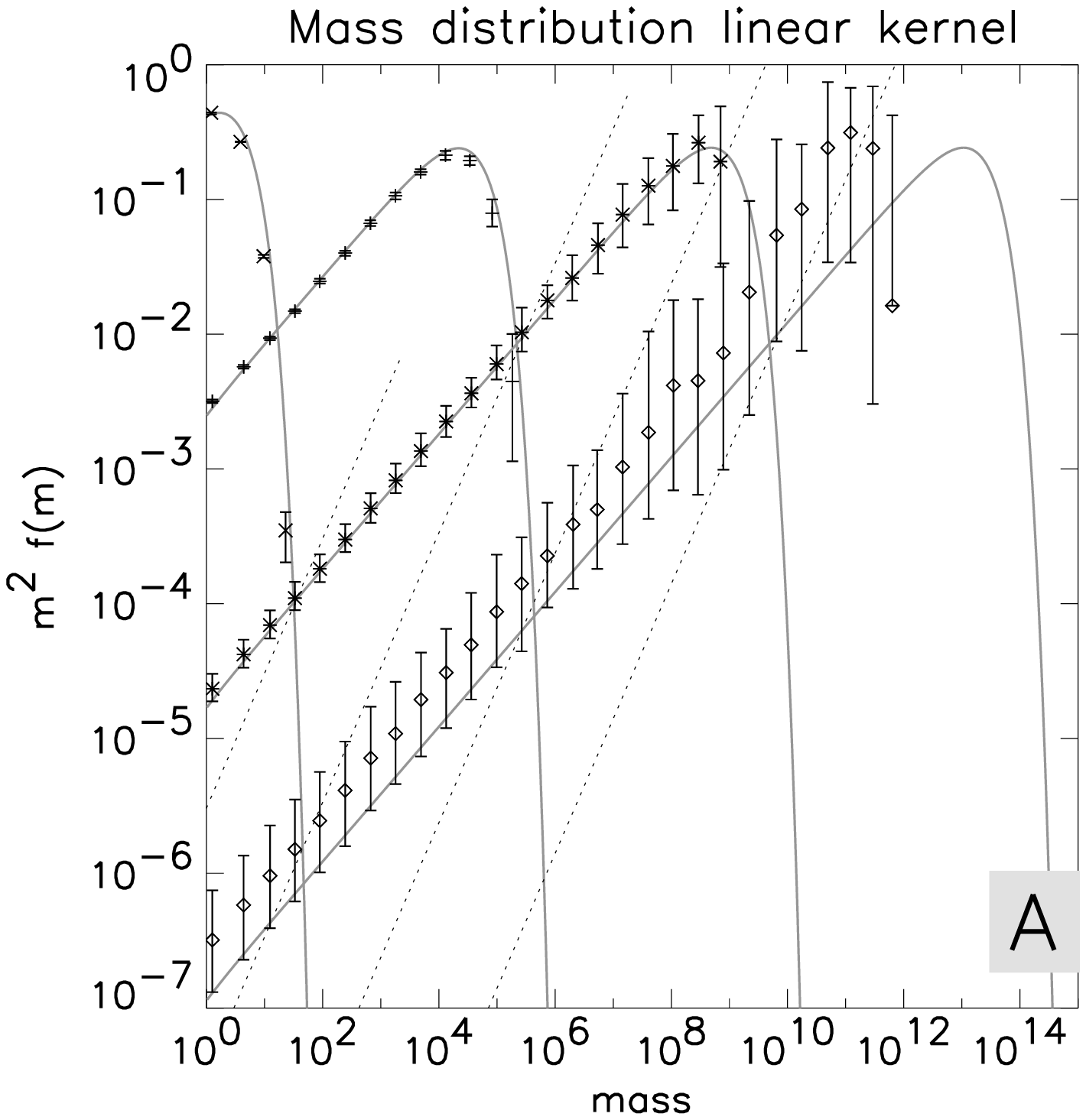}
    \includegraphics{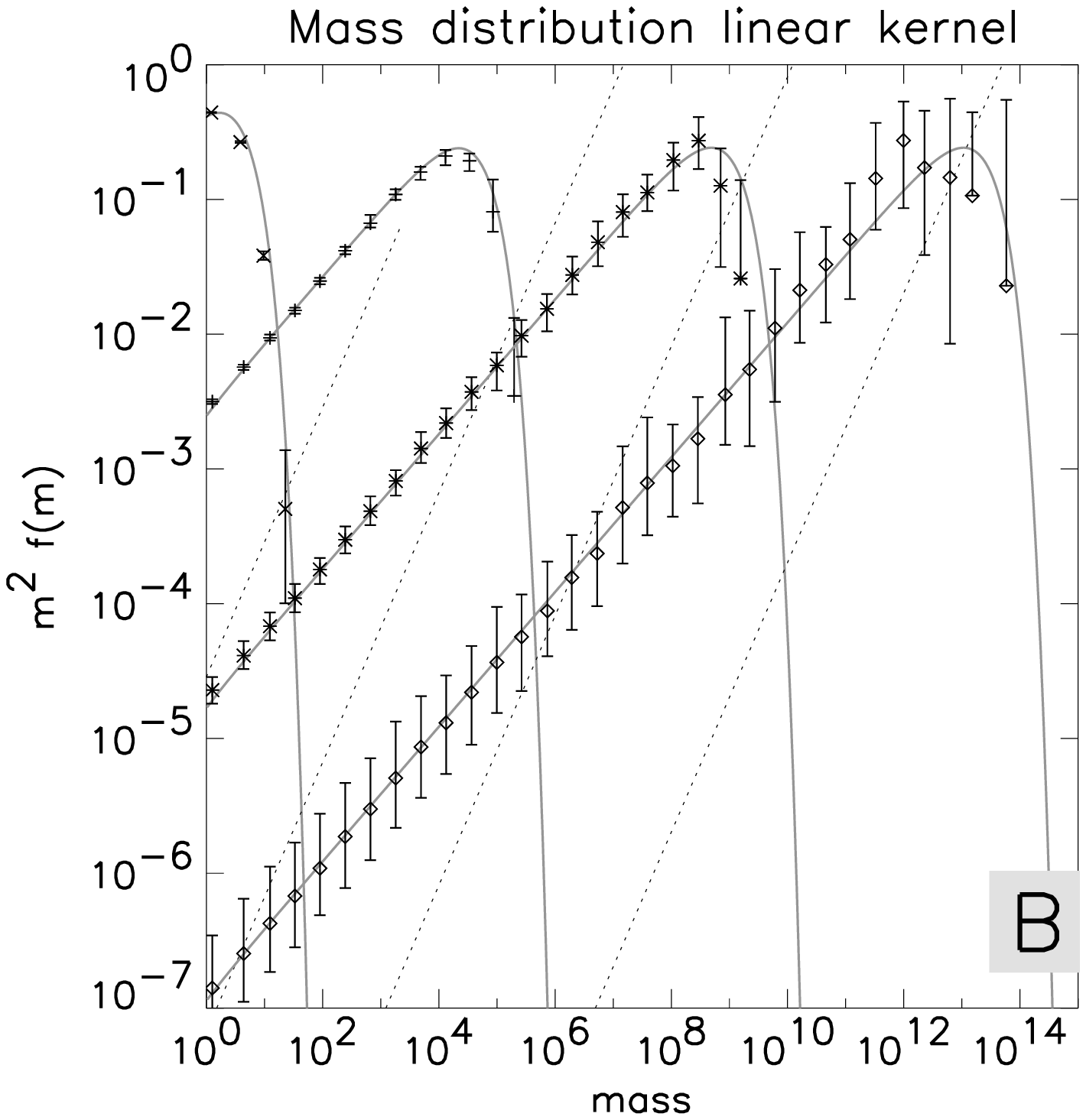}
  }
  \caption{Tests of the Monte Carlo coagulation code. Linear kernel $\left( K_{ij}= \frac{1}{2} (m_i+m_j) \right)$. -- (A) fixed $N$ ($N=200\,000$). The distribution function $f$ is computed at times $t=(1,10,20,30)$. At $t=30$ insufficient mass is present to provide a good fit to the analytical distribution. -- (B) fixed $N_\mathrm{s}$ . To obtain better correspondence with theory, the number of particles, $N$, is increased as the simulation progresses, such that more volume is sampled. $N_\textrm{s}$ is fixed at 20\,000 and $N$ ends with over a million particles. The average corresponds well to the theory, yet the amount of CPU time is disproportionally larger at the later times. }
  \label{fig:test3}
\end{figure*}

The Monte-Carlo coagulation model described above is tested with kernels that have an analytic solution. These are \textit{I}) the constant kernel, $K_{ij}=1$, and \textit{II}) the linear kernel, $K_{ij}= \frac{1}{2}(m_i+m_j)$. The evolution of the mean mass of the distribution, $\langle m \rangle$ for these kernels is\footnote{\changed{The mean mass of the population is inversely proportional to the number of particles per unit volume.}} \changed{ \citep{1979ApJ...229..242S,1994Icar..107..404T} }
\begin{equation}
    \langle m \rangle = \
    \begin{cases}
        m_0(1+\frac{1}{2}t)  & \qquad \textrm{constant kernel} \\
        m_0 \exp \left[ \frac{1}{2} t \right]  & \qquad \textrm{linear kernel},  \\
    \end{cases}
\end{equation}
where the distribution at $t=0$ is monodisperse of mass $m_0$.
Well-known coagulation models have either $K \propto m^{1/3}$ (Brownian coagulation) or $K\propto m^{2/3}$ (geometric area), but here, due to changes in $\psi$ and $\Delta \mathrm{v}$, we should be prepared for $K$ to vary with time. Thus, it is important that the Monte-Carlo code passes both these tests. Initial conditions for these test cases are monodisperse with all relevant parameters put at 1 at $t=0$ (i.e, $m_0=1$ and $f(t=0)=1$, $\rho_\mathrm{d}=1$, $\rho_\mathrm{s}=1$) and we do not take porosity into account ($\psi=1$ always). At various times the particles are binned by mass and the distribution function $f(m)$ is determined by summing over the masses in the bin and dividing by the width of the bin (to get the spectrum) and the volume of the simulation (to get the density). Multiple runs of the simulation then provide the spread in $f$. Figures \ref{fig:test1} and \ref{fig:test3} present the results. On the y-axis $f(m)$ is multiplied by $m^2$ to show the mass-density per logarithmic bin. Analytical solutions \changed{\citep{1994Icar..107..404T}} are overplotted by solid curves, while the dotted line shows the (hypothetical) distribution function if all the bins would be occupied by 1 particle. Thus, the dotted line corresponds to $m^2f(m) = m^2/ {\cal V} w_\mathrm{b} \approx m/{\cal V}$, because the widths of the bins, $w_\mathrm{b}$, are also exponentially distributed. In a single simulation, the distribution function should lie above this (auxiliary) line and the vertical distance to this line is a measure for the number of particles in a bin. 

Figure \ref{fig:test1} shows that the code passes the constant kernel test with flying colours. The spread in the data is limited, and it does not noticeably increase with time. The linear kernel (Fig.\ \ref{fig:test3}A), on the other hand, shows a different story. Here the mean mass, $\langle m \rangle$, as well as the \change{peak mass}, $m_\textrm{p}$ -- defined as the peak of the $m^2 f(m)$ size distribution -- evolve exponentially with time. \change{Note that the position of $m_\textrm{p}$ only depends on the particles that contain most of the mass, while $\langle m \rangle$ is also sensitive to the total number of particles. Therefore, $\langle m \rangle$ lags $m_\textrm{p}$ at any time and one can show that the gap between the two also increases exponentially with time.} Inevitably, at some point in time, the theoretical value of the $m^2f(m)$ mass-peak becomes larger than the total mass present inside ${\cal V}$. This corresponds to the crossing of the dotted line at, e.g., $t\simeq 20$ in Fig.\ \ref{fig:test3}A. In other words, the duplication mechanism, needed to conserve $N$ but which has the additional effect of enlarging ${\cal V}$, is incapable of keeping up with the exponential growth: `runaway particles' could have formed, but the simulated volume ${\cal V}$ is just not large enough to take them into account. The postulate of the `duplication mechanism' -- the particle distribution evolves similarly in and outside ${\cal V}$ -- then breaks down. The only way to avoid this effect is to enlarge ${\cal V}$ by having more particles in the simulation, i.e., to improve the `numerical resolution'. In Fig.\ \ref{fig:test3}B, we show the results, in which $N_\textrm{s}$, instead of $N$, is held constant (Sect.\ \ref{sec:mcimplement}). In these simulations $N$ increases with time, starting with $N=20\,000$ and ends with more than a million particles. The distribution now represents more closely the theoretical curve. Growth factors of 14 orders of magnitude in mass ($\simeq 5$ in size) can then be accurately simulated.  

The drawback, however, is that the computation slows down as $N$ increases, since the relative increase in the average mass is inversely proportional to $N$, i.e., $\Delta \langle m \rangle/\langle m \rangle \propto N^{-1}$. \changed{These simulations therefore take much more CPU time. It is clear that a fundamental limit is reached, in which, given a certain CPU power, the calculation of the mass-distributions can only be achieved for a limited range. A way to overcome this problem is to collide multiple particles per event. In such an algorithm collisions are no longer between two particles but between two groups of particles. Although this approximates the collisional process, the coagulation can be speeded up by grouping especially small particles into a single unit. We will discuss the grouping algorithm and its implication in a future paper. For the simulations in Sect.\ \ref{sec:res}, the product of $N$ and $N_\mathrm{s}$ has been kept constant, which ensures a} constant growth (in exponential terms) per cycle. We have fixed $\sqrt{N\times N_\mathrm{s}}$ at 20\,000 but made sure that the numerical resolution issues \changed{as discussed here for the linear kernel, did not occur. Fortunately, realistic mass-distributions are not that broad as in the analytic, linear kernel; e.g., the smaller particles are quickly removed due to Brownian coagulation.}  

In summary, we have built an efficient Monte Carlo code to follow coagulation. The advantage of this code (above other numerical methods) is that it is intuitive, simple to implement and expand, and that it takes full account of the stochastic nature of coagulation. Minor disadvantages are the $N^2$ dependence on CPU time, the somewhat artificial duplication procedure, and the resolution problems shown for the linear kernel at high $m$ as the simulation progresses. We have addressed these concerns and \changed{developed methods} suitable for the scope of this work. DSMC-coagulation methods are very appropriate to work in conjunction with multi-parameter models. The collision model of Sect.\ \ref{sec:model}, with mass, $m$, and \changed{enlargement parameter, $\psi$, as variables}, can now be put into an evolutionary setting.

\section{Results\label{sec:res}}
\subsection{Application to a protoplanetary disk}
The collision model of Sect.\ \ref{sec:model} is embedded in the Monte Carlo code of the previous section and applied to the protoplanetary disk (Sect.\ \ref{sec:environ}). The coagulation is restricted to a turbulent environment in which particles are well mixed with the gas, yet do not fragment upon collision. For these reasons, we start out with a monodisperse population of sub-micron-sized grains (we choose them to be $a_0=0.1\ \mu\mathrm{m}$) present at $z=H_g$, i.e., at the scaleheight of the disk, where the density is a factor $\exp [-0.5]$ lower than at the mid-plane. Its collisional evolution within the gaseous nebula is followed until the point that systematic motions, i.e., settling to the mid-plane, start to dominate. Thus, particles are either present and well mixed by turbulence, or have started to settle and are no longer in the region of interest. Settling is then modelled as a sudden phenomenon. The reality is, of course, a more gradual transition, but the discrete picture here is not a bad approximation, since the vertical structure is quickly established \citep{2004ApJ...601.1109Y}. Settling occurs at the point when the friction time of a particle has exceeded a critical friction time, $\tau_\mathrm{rain}$, such that the scaleheight of the particle, $h(\tau_\mathrm{f})$, becomes smaller than that of the gas, i.e., $h(\tau_\textrm{f}) < H_\mathrm{g}$. If self-gravity is neglected (valid in the gaseous nebula) and Schmidt numbers, Sc, are close to unity,\footnote{The Schmidt number measures the ratio of the gas to particle diffusivity; it is supposed to be close to unity if $\tau_\mathrm{f} < t_0$. \citep{2004ApJ...614..960S}} then $h$ can be obtained by equating the particle diffusion timescale, $t_\mathrm{diff} = h^2/\nu_\mathrm{T}$, with the particle settling timescale, $t_\mathrm{settl}=(\Omega^2 \tau_\mathrm{f})^{-1}$, i.e., $h(\tau_\mathrm{f}) = H_\mathrm{g} \sqrt{\alpha_\mathrm{T} / \Omega \tau_\mathrm{f} }$ \citep[cf.][]{2004ApJ...614..960S,2005astroph0508659}. The critical friction time is then
\begin{equation}
  \tau_\mathrm{f}  = \frac{\rho_\mathrm{s}}{c_\mathrm{g} \rho_\mathrm{g}} \frac{a^*}{\psi^{2/3}} = \tau_\mathrm{rain} = \frac{\alpha_\mathrm{T}}{\Omega},
  \label{eq:critfric}
\end{equation}
with $a^*$ the compact size of particles (see Fig.\ \ref{fig:fractalfig}). As expected, higher values of $\alpha_\mathrm{T}$ and higher gas densities (lower $\tau_\mathrm{f}$) delay the onset of settling in the sense that the particle has to grow further in size before it arrives at the critical friction time. Alternatively, increasing $\psi$ also delays settling.

In the code, settling is implemented as a `rain-out': the particle is removed from the simulation and the spatial dust density decreases. The evolution of these `rain-out particles' during settling is not traced anymore. The focus stays on the particles that remain in the layers above the mid-plane. Their evolution is followed for $10^7\ \mathrm{yr}$.

The $\alpha_\mathrm{T}$ parameter is one of the major uncertainties concerning the characterisation of protoplanetary disks. One of the prime candidates for turbulence is the magneto-rotational instability \citep{1991ApJ...376..223H,1991ApJ...376..214B}, which seems to be most robust in well-ionised regions, i.e., in the upper layers of the disk. Another way to characterise $\alpha_\textrm{T}$ is to relate it to the observed accretion rate, $dM/dt$, \citep{2005ASPC..341..732C} and then values in the range of $10^{-4} \lesssim \alpha_\textrm{T} \lesssim 10^{-2}$ seem plausible. Yet, despite its uncertainty, $\alpha_\textrm{T}$ appears in key expressions as, e.g., $\Delta \mathrm{v}_{ij}$ and $\tau_\mathrm{rain}$. Therefore, models are run that cover a large range in $\alpha_\mathrm{T}$: $\alpha_\mathrm{T} = 10^{-6}-10^{-2}$. Furthermore, we divide the runs in two categories, which reflect the spatial position in the solar nebula. The `inner' models correspond to conditions at 1 AU where the monomers are quartz with internal density $\rho_\mathrm{s} = 3.0\ \mathrm{g\ cm}^{-3}$ and surface energy density of $\gamma = 25\ \mathrm{ergs\ cm}^{-2}$. The `outer' models correspond to conditions at 5 AU, where the coagulation is that of ices ($\rho_\mathrm{s} = 1\ \mathrm{g\ cm}^{-3}$ and $\gamma = 300\ \mathrm{ergs\ cm}^{-2}$) and with an enhanced surface density of a factor 4.2 over the minimum solar nebula \citep{1986Icar...67..375N}. For comparison, we also run compact models for the $\alpha_\mathrm{T}=10^{-4}$ runs. Compact models (denoted by the C-suffix in Table\ \ref{tab:runs}) are models where the internal structure does not evolve, i.e., $\psi=1$ by definition. An overview of all the models is given in Table \ref{tab:runs}.
\begin{table}
    \caption{Overview of all the runs}
    \label{tab:runs}
    \centering
    \begin{tabular}{lllll}
        \hline
        \hline
        Model-id$^a$&   \multicolumn{3}{c}{Model Parameters}        & Notes \\
                    &   $R\ ^b$ & $\alpha_\mathrm{T}$   & $\delta$  &       \\
        \hline
        R1Ta4-P     &   1 AU     & $10^{-4}$             & 0.95      &  default model\\
        R1Ta2-P     &   1 AU     & $10^{-2}$             & 0.95      &  increased turbulence \\
        R1Ta6-P     &   1 AU     & $10^{-6}$             & 0.95      &  decreased turbulence \\
        R1Ta4-C     &   1 AU     & $10^{-4}$             & 2/3       &  compact model \\
        R5Ta4-P     &   5 AU     & $10^{-4}$             & 0.95      &  default model at 5AU \\
        R5Ta2-P     &   5 AU     & $10^{-2}$             & 0.95      &  increased turbulence \\
        R5Ta6-P     &   5 AU     & $10^{-4}$             & 0.95      &  decreased turbulence \\
        R5Ta4-C     &   5 AU     & $10^{-6}$             & 2/3       &  compact model \\
        \hline
        \hline
    \end{tabular}\\
    \begin{tabular}{r@{\hspace{0.5mm}}p{8cm}}
        $^a$ &Model names are intended to be mnemonic. `Rx' stands for radius at x AU. Ta, denotes the strength of the turbulence, e.g., Ta4 stands for $\alpha_\mathrm{T} = 10^{-4}$. Finally, the suffix denotes whether models are porous (P) or compact (C).\\
        $^b$ &At 1 AU quartz particles coagulate: $\rho_\mathrm{g}/\rho_\mathrm{d} = 240$, $\gamma=25\ \mathrm{ergs\ cm}^{-2}$, $\rho_\mathrm{s}=3.0\ \mathrm{g\ cm}^{-3}$; at 5 AU coagulation is between ices: $\gamma = 370\ \mathrm{ergs\ cm^{-2}}$, $\rho_\mathrm{s}=1.0\ \mathrm{g\ cm^{-3}}$, $\rho_\mathrm{g}/\rho_\mathrm{d} = 57$. The gas parameters correspond to a minimum mass solar nebula model (see Sect.\ \ref{sec:environ}).\\
    \end{tabular}
\end{table}

\subsection{Particle growth and compaction}
\begin{figure*}
  \resizebox{12cm}{!}{\includegraphics{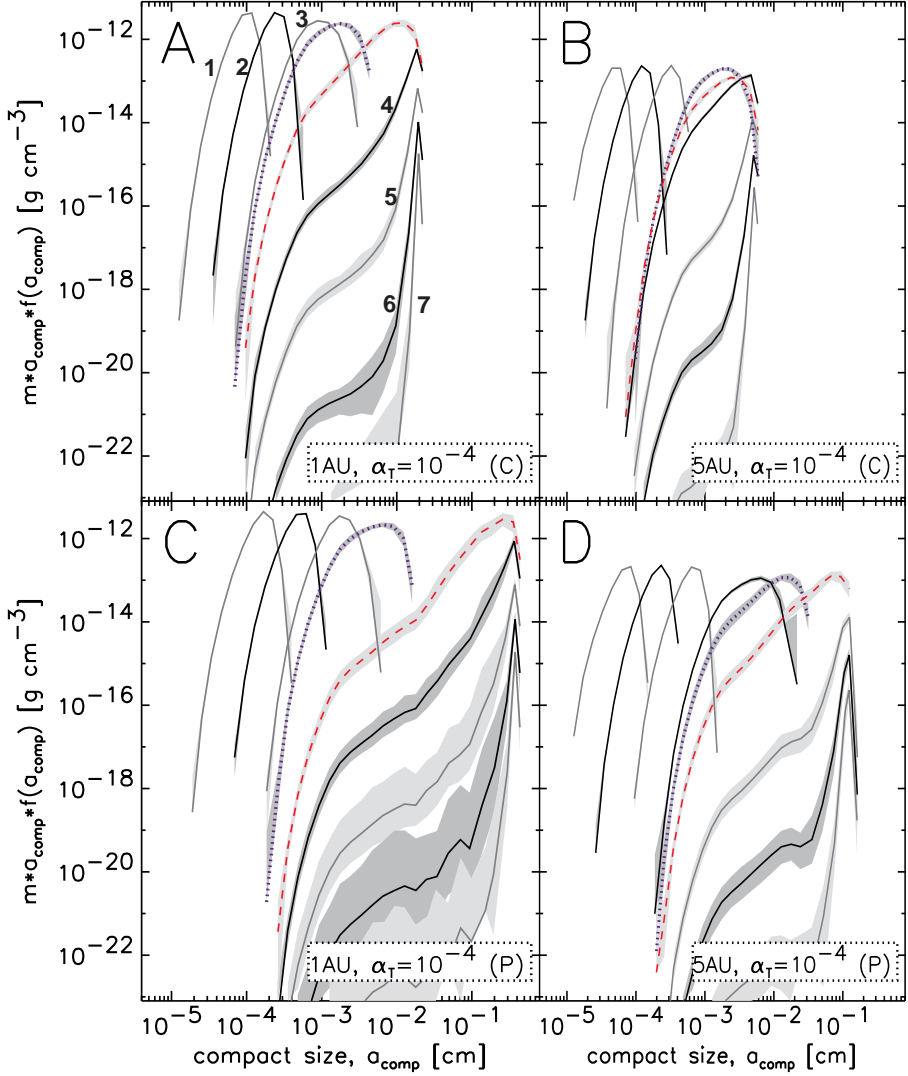}}
\caption{The mass function plotted at various times for the $\alpha_\mathrm{T}=10^{-4}$ models. The panels compare the coagulation of the compact models ($\psi=1$, top panels) with those where porosity effects are included (bottom panels). Left (right) panels show the coagulation of quartz (ice) particles at 1 AU (5 AU).  Each plot shows the mass function at every logarithmic interval in time from $t=10\ \mathrm{yr}$ until $t = 10^7\ \mathrm{yr}$. In the first $\sim 10^2\ \mathrm{yr}$ Brownian motion dominates the coagulation. Subsequent evolution is driven by turbulence-induced velocity differences and includes the moment of first compaction (blue, dotted line) and first rain-out (red, dashed line). After rain-out ($t \gtrsim 10^4\ \mathrm{yr}$), the mass density in the gaseous nebula decreases and the mass function collapses. In the compact models the blue, dotted curve also corresponds to the first time that $E > E_\mathrm{roll}$. Greyscales indicate the spread in the 50 realisations of the simulation.}
  \label{fig:massfunc1}
\end{figure*}

\begin{figure*}[t]
  \includegraphics[clip=true,scale=0.55]{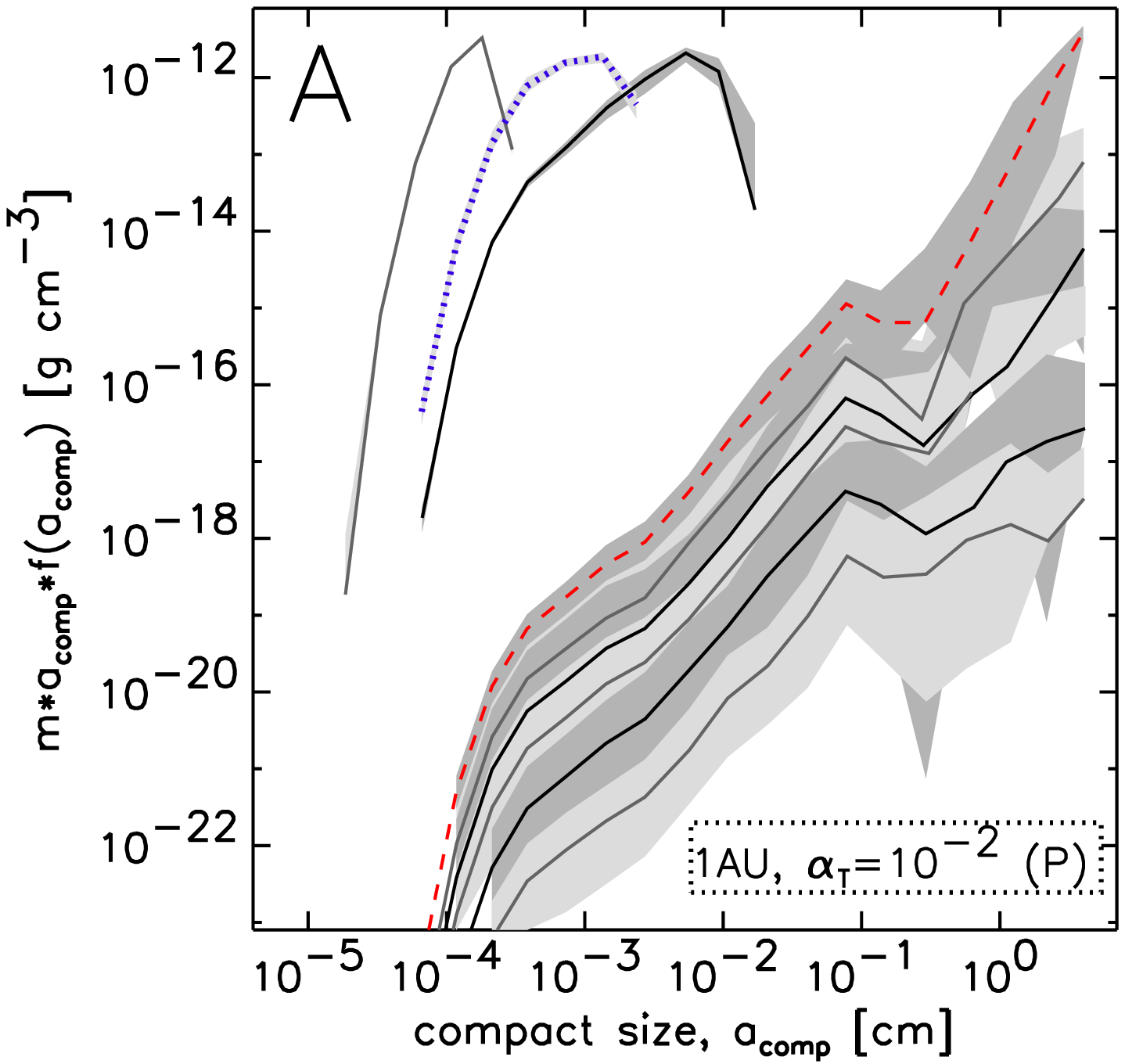}\includegraphics[clip=true,scale=0.55]{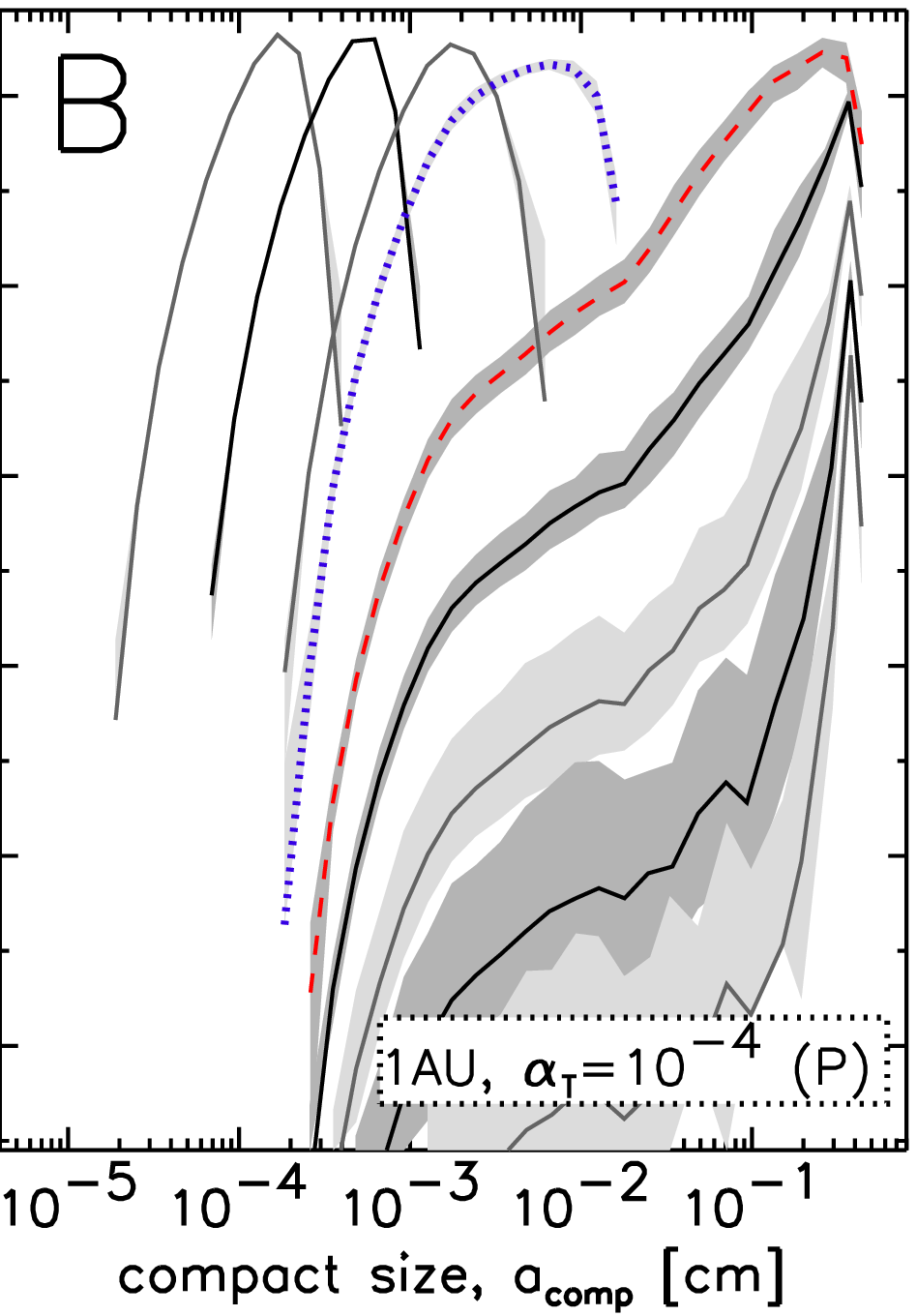}\includegraphics[clip=true,scale=0.55]{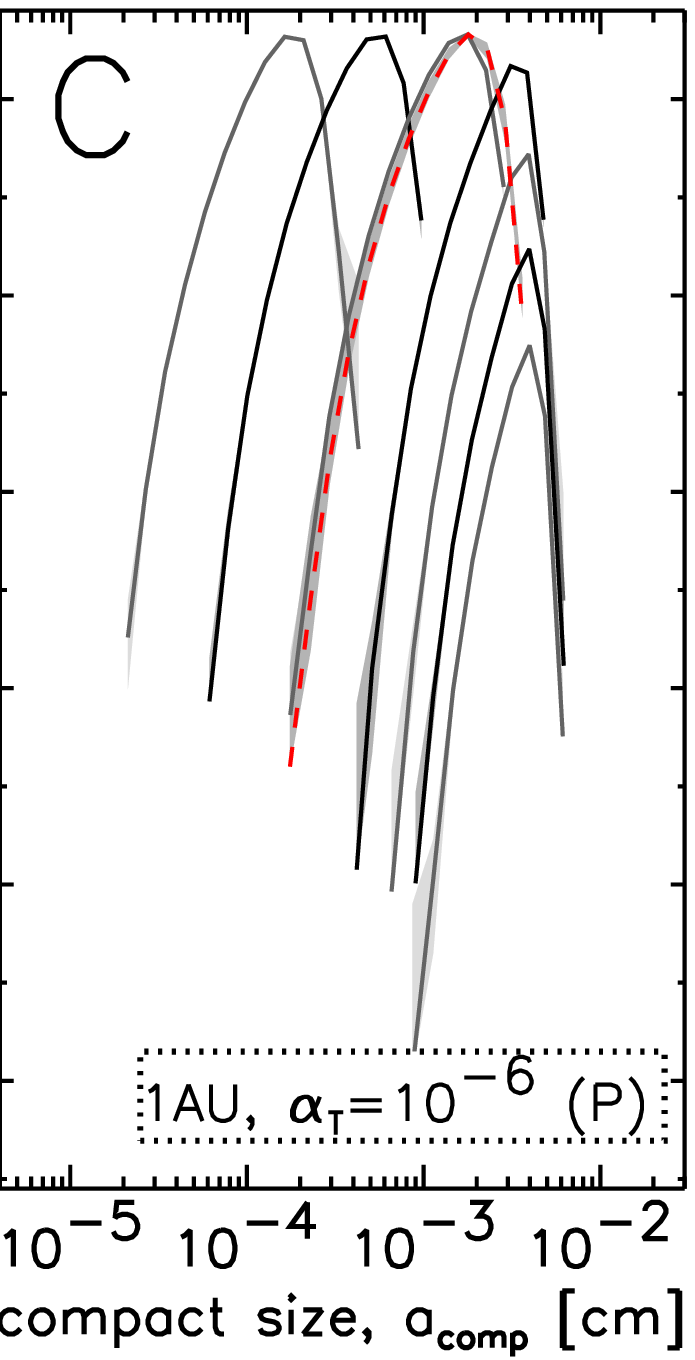}
  \caption{Effects of turbulence on the coagulation. Panels show the collisional evolution at 1 AU of the porous models, yet with $\alpha_\mathrm{T}$ values of $10^{-2}$ (A), $10^{-4}$ (B) and $10^{-6}$ (C). The scaling of the axis is the same throughout the panels. In the $\alpha_\mathrm{T}=10^{-2}$ models the spread in the runs is very large, causing the error bars to overlap. In the $\alpha_\mathrm{T}=10^{-6}$ model the particles rain-out without compacting.}
  \label{fig:massfunc2}
\end{figure*}

\begin{figure}
  \resizebox{\hsize}{!}{
    \includegraphics[width=88mm]{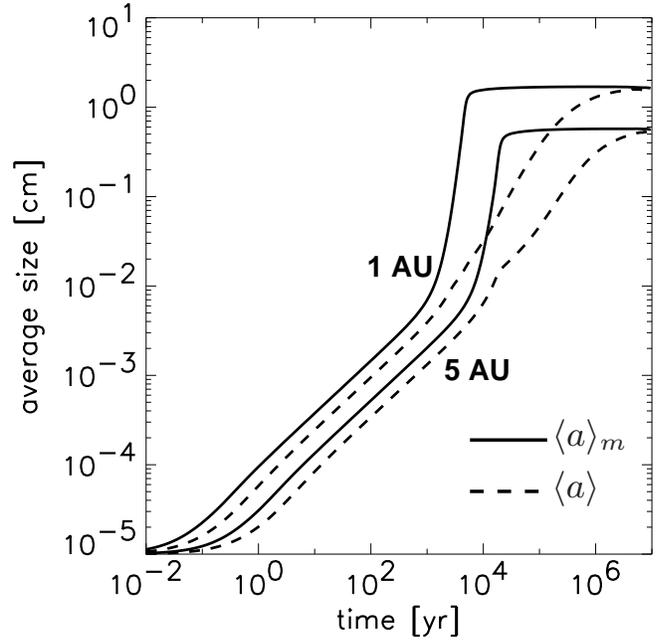}
    }
  \caption{Evolution of the size distribution with time. The mass-weighted size, $\langle a \rangle_m$ (solid line), and the mean size, $\langle a \rangle$ (dashed line), are calculated for the default models ($\alpha_\mathrm{T} = 10^{-4}$) at 1 AU (top lines) and 5 AU (bottom lines). Lines are averaged over the 50 simulation runs. After $t \gtrsim 10^3\ \mathrm{yr}$ the coagulation drives most of the mass into the largest particles.}
  \label{fig:evoltime}
\end{figure}

In Fig.\ \ref{fig:massfunc1} mass distributions are shown at various times during their collisional evolution. On the y-axis the mass function is plotted in terms of $m\cdot a^\ast \cdot f(a^*)$, which shows the mass-density, i.e., the mass of grains of compact size $a^*$ in logarithmic bins. The panels compare the results of compact coagulation (panels A and B) with those of porous coagulation (panels C and D) for a turbulent strength parameter of $\alpha_\mathrm{T}=10^{-4}$. The coagulation is calculated at 1 AU (quartz; panels A and C) and at 5 AU (ices, panels B and D). In Fig.\ \ref{fig:massfunc2} the $\alpha_\mathrm{T}=10^{-4}$ model at 1 AU is compared to other $\alpha_\mathrm{T}$ models at 1 AU (see Table\ \ref{tab:runs}). In Fig.\ \ref{fig:evoltime} averages of the size distributions of Figs.\ \ref{fig:massfunc1}A, C are shown explicitly with time. Here $\langle a \rangle_m$ is the mass-weighted size,
\begin{equation}
    \langle a \rangle_m = \frac{1}{\sum_i m_i} \sum_i m_i a_i,
    \label{eq:mws}
\end{equation}
of the population. Thus, while $\langle a \rangle$ gives the average particle size, $\langle a \rangle_m$ corresponds to the (average) size to which a unit of mass is confined. 
Because of this weighting, $\langle a \rangle_m \ge \langle a \rangle$, with the equality valid only for monodisperse distributions.

\begin{figure*}
  \resizebox{\hsize}{!}{
    \includegraphics[clip=true,scale=0.5]{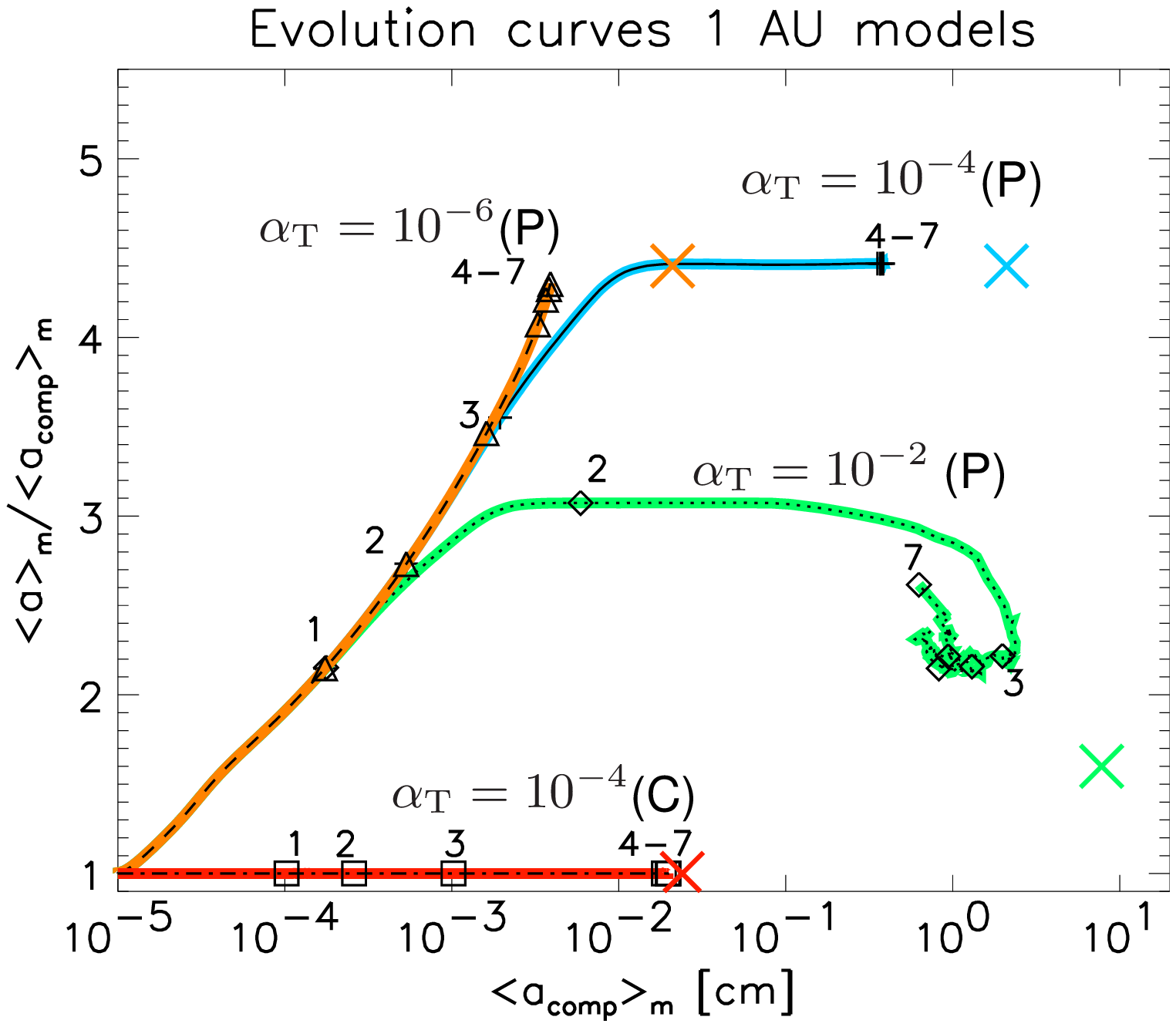}\includegraphics[clip=true,scale=0.5]{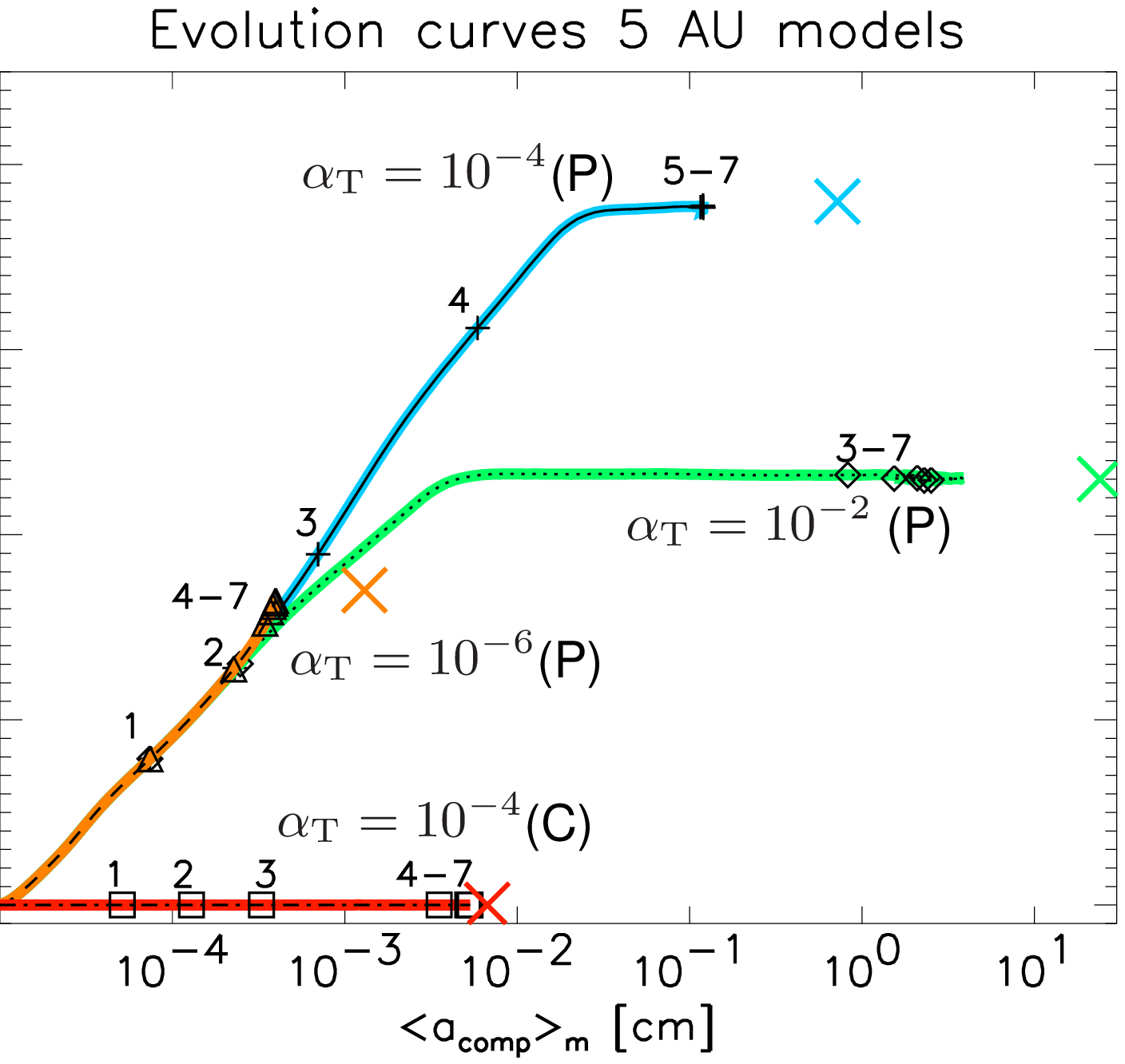}
  }
  \caption{Evolution curves of various models. In these panels the \changed{enlargement factor} is characterised by the ratio of the mass-weighted porous size, $\langle a \rangle_m$, over the mass-weighted compact size, $\langle a_\mathrm{comp} \rangle_m$, ($\langle a^* \rangle_m$ in the text), against $\langle a_\textrm{comp} \rangle_m$. Rising curves correspond to an increase of porosity due to fractal growth, and horizontal or declining lines indicate compaction. Numbers give the temporal stage of the coagulation (i.e., $t=10^i$). The detached, coloured crosses indicate the average \textit{porous sizes} (x-axis) and the size enhancement of the rain-out particles (y-axis).}
  \label{fig:evolcurves}
\end{figure*}
The porosity evolution is also displayed in Fig.\ \ref{fig:evolcurves}, where we plot the ratio of $\langle a \rangle_m$ to $\langle a^* \rangle_m$. $\langle a^* \rangle_m$ is defined analogously to Eq.\ (\ref{eq:mws}), but then for the compact particle size. $\langle a \rangle_m / \langle a^* \rangle_m$ then gives the \changed{mass-weighted averaged enhancement of the geometrical radius of the particles (see, Fig.\ \ref{fig:fractalfig}) . In the case of monodisperse populations or runaway growth one size dominates the average and \change{the ratio is directly related to $\langle \psi \rangle_m$ as} $\langle a \rangle_m / \langle a \rangle_m = \langle \psi \rangle_m^{1/3}$} This ratio is plotted vs.\ $\langle a^* \rangle_m$ in Figs.\ \ref{fig:evolcurves} for various models. It shows the build-up of porosity during the fractal stage, the stabilisation during the compaction stage, upto the stage where the particles rain-out. In Fig.\ \ref{fig:evolcurves} the average sizes of some of these rain-out particles are indicated by the detached crosses. Note that (for illustrative reasons) the x-axis for these particles corresponds to porous size, $a$, rather than compact size, $a^*$. Properties of the `rain-out' population are given in Table \ref{tab:rainout}. In Fig.\ \ref{fig:evolcurves} the temporal stage is indicated by numbered ticks.
\begin{table*}[tbp]
  \caption{Properties of the particle distribution at rain-out.}
  \label{tab:rainout}
  \centering
  \begin{tabular}{lllllll}
    \hline
    \hline
    model     &   $\langle a \rangle$   & $\langle m \rangle $          & \changed{$\langle \psi \rangle$} & $\langle a/a^* \rangle$ & $\langle t_\mathrm{rain} \rangle$     & $\langle t_\mathrm{99\%} \rangle$     \\
              &   cm                    & gr                            &                         &                         & yr                                    & yr                                    \\
    (1)       &   (2)                   & (3)                           & \changed{(4)}           & (5)                     & (6)                                   & (7)               \\
    \hline                                                        
    R1Ta4-P   & $ 2.1\pm0.1$            & $1.4\pm0.1$                   & $(8.6\pm0.2)\times10^1$ & $4.4\pm0.0$             & $(4.7\pm 0.2)\times 10^3$             & $(4.5\pm 1.0)\times 10^4$     \\
    R1Ta4-C   & $ 2.4\times 10^{-2}$    & $(1.9\pm0.2)\times 10^{-4}$   & 1                       & 1                      & $(2.6\pm 0.1)\times 10^3$             & $(4.4\pm 0.3)\times 10^4$     \\
    R5Ta4-P   & $ 0.72\pm0.04 $         & $(1.4\pm0.1)\times 10^{-2}$   & $(1.1\pm0.1)\times10^2$ & $4.8\pm0.1$             & $(1.9\pm 0.1)\times 10^4$             & $(2.2\pm 0.3)\times 10^5$     \\
    R5Ta4-C   & $ 6.7\times 10^{-3} $   & $1.2\times 10^{-6}$           & 1                       & 1                       & $(8.1\pm 0.3)\times 10^3$             & $(2.1\pm 0.1)\times 10^5$     \\
    R1Ta2-P   & $ 7.8\pm 0.7$           & $(1.6\pm0.3)\times 10^3$      & $(3.9\pm1.1)$           & $1.6\pm 0.1$            & $(2.6\pm 0.3)\times 10^2$             & $(4.0\pm 19)\times 10^5\ ^{b}$     \\
    R5Ta2-P   & $ 24 \pm 2$             & $(1.7\pm0.3)\times 10^3$      & $(3.6\pm0.1)\times10^1$ & $3.3$                   & $(1.4\pm 0.1)\times 10^3$             & $(4.5\pm 4.6)\times 10^3\ ^{b}$     \\
    R1Ta6-P   & $ 2.1\times 10^{-2}$    & $1.3\times 10^{-6}$           & $(8.7\pm0.1)\times10^1$ & $4.4$                   & $(1.3\pm 0.1)\times 10^3$             & $2.3\times 10^6$     \\
    R5Ta6-P   & $ 1.3\times 10^{-3}$    & $(4.6\pm0.1)\times 10^{-10}$  & $(2.0\pm0.0)\times10^1$ & $2.7$                   & $61\pm 7.2$                           & $1.1\times 10^5$     \\
    \hline
  \end{tabular}\\
  \begin{tabular}{lp{12cm}}
    $^a$  & Entries denote: (1) model-id (see Table\ \ref{tab:runs}); (2) size at rain-out; (3) mass at rain-out; \changed{(4) enlargement factor at rain-out;} (5) size enhancement at rain-out; (6) time at which first rain-out occurs; (7) time-interval over which 99\% of the mass has rained-out. Values have been averaged over the 50 runs with the error bars reflecting the variations between the 50 runs of the simulation. (The spread is only given when the rms-value exceeds the second significant digit.) \\
    $^b$  & Some simulations did not achieve a 99\% rain-out of the density, so that $t_\mathrm{99\%} > 10^7\ \mathrm{yr}$. This caused the large spread.\\
  \end{tabular}
\end{table*}

The evolution of the mass-distribution (Figs.\ \ref{fig:massfunc1}, \ref{fig:massfunc2}, \ref{fig:evoltime}) can be divided into three stages. Initially, since particles start out as grains with sizes of $0.1\ \mu\mathrm{m}$, Brownian motion dominates. The size-distribution is therefore rather narrow, because the Brownian collision kernel favours the lighter particles. Quickly ($\sim 10^2\ \mathrm{yr}$), however, turbulent velocities become dominant and relative velocities are now highest for the more massive (high $\tau_\mathrm{f}$) particles. Once the first compaction event occurs (dotted, blue line\footnote{References to colours only apply to the electronic version of this paper.}) $\tau_\mathrm{f}$ enters the regime in which it becomes (at least) proportional to size and the pace of coagulation strongly accelerates toward larger sizes. \changed{These findings correspond well with the simple analytical model of \citet{2004ASPC..309..369B}, where in his Fig.\ 5 the Brownian motion driven growth is also followed by a stage in which the growth is exponential.} This evolution could be called `run-away', were it not for the fact that \changed{(in our case)} the particle distribution is truncated at $\tau_\mathrm{rain}$ (Eq.\ \ref{eq:critfric}). The distribution at the first rain-out event is shown by the dashed, red line. Thereafter, the mass function collapses and evolves to a monodisperse population, close to the rain-out size (Figs.\ \ref{fig:massfunc1},\ref{fig:evoltime}). Two causes conspire to make these particles favoured: first, large particles can only be (efficiently) removed by a collision with a similar-sized particle (and no longer by a larger particle since these have disappeared); second, in the $\alpha_\mathrm{T}=10^{-4}$ models friction times are always in the $\tau_f < t_s$ regime and relative velocities between similar-sized particles are suppressed. 
For these reasons, particles in the $\alpha_\mathrm{T} = 10^{-4}$ models near rain-out deplete the smaller particles faster than they deplete themselves, and the size distribution evolves again to monodispersity. Note, however, that this behaviour is essentially caused by the imposed presence of a sharp cut-off size due to the rain-out criterion. In reality, a more smooth transition can be expected.

Although the qualitative trends between the porous and compact models are essentially the same -- fractal growth, compaction and run-away growth, rain-out and depletion -- it is unambiguously clear that the porous models evolve to larger particles, as is also seen in Table\ \ref{tab:rainout} in which the values for the rain-out particles are tabulated. The size difference at rain-out is $\sim 2$ order of magnitude in size and $\sim$ 4 orders of magnitude in mass. Particles only rain-out at $\tau_\mathrm{rain}$ and in the porous models particles have to grow much further before the critical friction time is reached (Eq.\ \ref{eq:critfric}). The inclusion of porosity in the coagulation models thus allows particles (when $\alpha_\mathrm{T} > 10^{-4}$) to grow to cm/dm sizes in the gaseous nebula, i.e., before settling to the mid-plane. 

Apart from determining the size at which particles rain-out, $\alpha_\mathrm{T}$ also determines the pace of coagulation. In the $\alpha_\mathrm{T}=10^{-2}$ models (Fig.\ \ref{fig:massfunc2}A) coagulation is rapid. Also, a large degree of stochasticity is seen among different models. In the $\alpha_\mathrm{T}=10^{-6}$ models, on the other hand, the turbulent velocities are very small and the support against gravity is weak, such that that rain-out happens before any compaction takes place. These models are most reminiscent of the `laminar nebula', where systematic (i.e., settling) velocities dominate and are therefore most prone to gravitational instability effects \citep{2005astro.ph..9701H}. The comparison between the various $\alpha_\mathrm{T}$-models is perhaps best seen in the `evolution tracks' of Figs.\ \ref{fig:evolcurves}. They show that initially all porous models follow the same (fractal) curve, until the moment compaction occurs. In the 1 AU, $\alpha_\textrm{T}=10^{-2}$ model a significant compaction of aggregates can clearly be observed (descending line). After $t > 10^3\ \textrm{yr}$, this is followed by a slight increase in $\langle a \rangle_m / \langle a_\textrm{comp} \rangle_m$; apparently due to the heavy rain-out, most collisions are again in the fractal regime. 

Comparing the coagulation of the two materials studied here, i.e., quartz for the 1 AU models and ice for the 5 AU models, one sees similarities in their collisional evolution (see also Fig.\ \ref{fig:evoltime}). It seems, however, that the coagulation at 5 AU is somewhat slower. This can be explained naturally because of the lower density, but also perhaps because compaction is more difficult to achieve due to the higher surface density ($\gamma$) of ices. Although the differences are subtle, one can see, e.g., in Fig.\ \ref{fig:evolcurves} that the curves of the $\alpha_\mathrm{T} = 10^{-4}, 10^{-2}$ porous models level-off at higher $\langle a \rangle_m$-to-$\langle a \rangle$ ratio than the corresponding 1 AU curves, indicating compaction is achieved `easier' at 1 AU. Also, at $\alpha_\mathrm{T} = 10^{-2}$ the 1 AU particles that rain-out are strongly compacted (an even higher $\alpha_\mathrm{T}$ would have led to fragmentation), while the rain-out particles at 5 AU do not compact considerably before rain-out (see also Table\ \ref{tab:rainout}). Thus, the larger surface energy ($\gamma$) of ices translates into a higher rolling energy and, subsequently, to decreased compaction.

\section{\label{sec:discuss}Discussion}
\begin{figure}
  \resizebox{\hsize}{!}{
    \includegraphics[width=88mm]{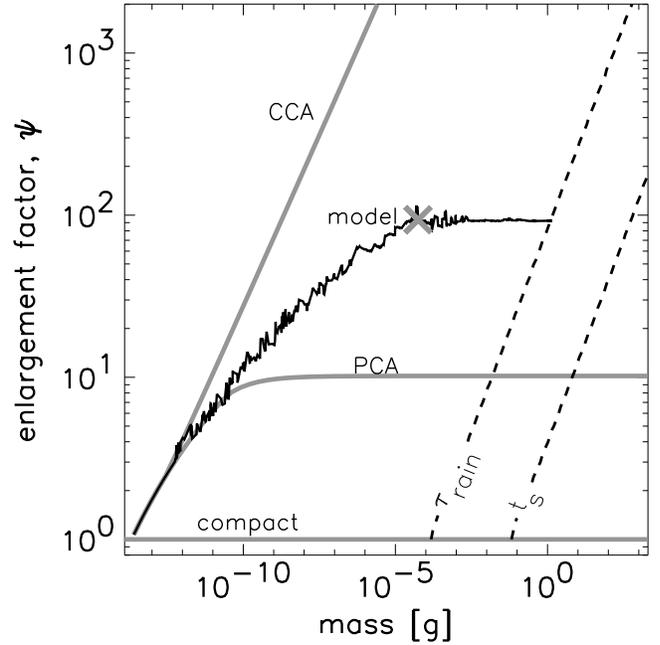}
  }
  \caption{The $m$-\changed{$\psi$} relation for several aggregation models. Plotted is $\psi(m)$ for: the most massive particle in one of the R1Ta4-P models (black curve); the PCA, CCA aggregation models, discussed in Sect.\ \ref{sec:ccapcacomp}, (grey curves); and compact ($\psi=1$) models. The dashed lines indicate points of equal friction times, $\tau_\mathrm{rain} \approx 500\ \mathrm{s}$ and $t_\mathrm{s}\approx 1\,600\ \mathrm{s}$. The cross shows the point at which the model experiences the first compaction event.}
  \label{fig:growthmodels}
\end{figure}
\begin{figure*}[t]
  \resizebox{\hsize}{!}{
    \includegraphics[scale=0.5, clip=true]{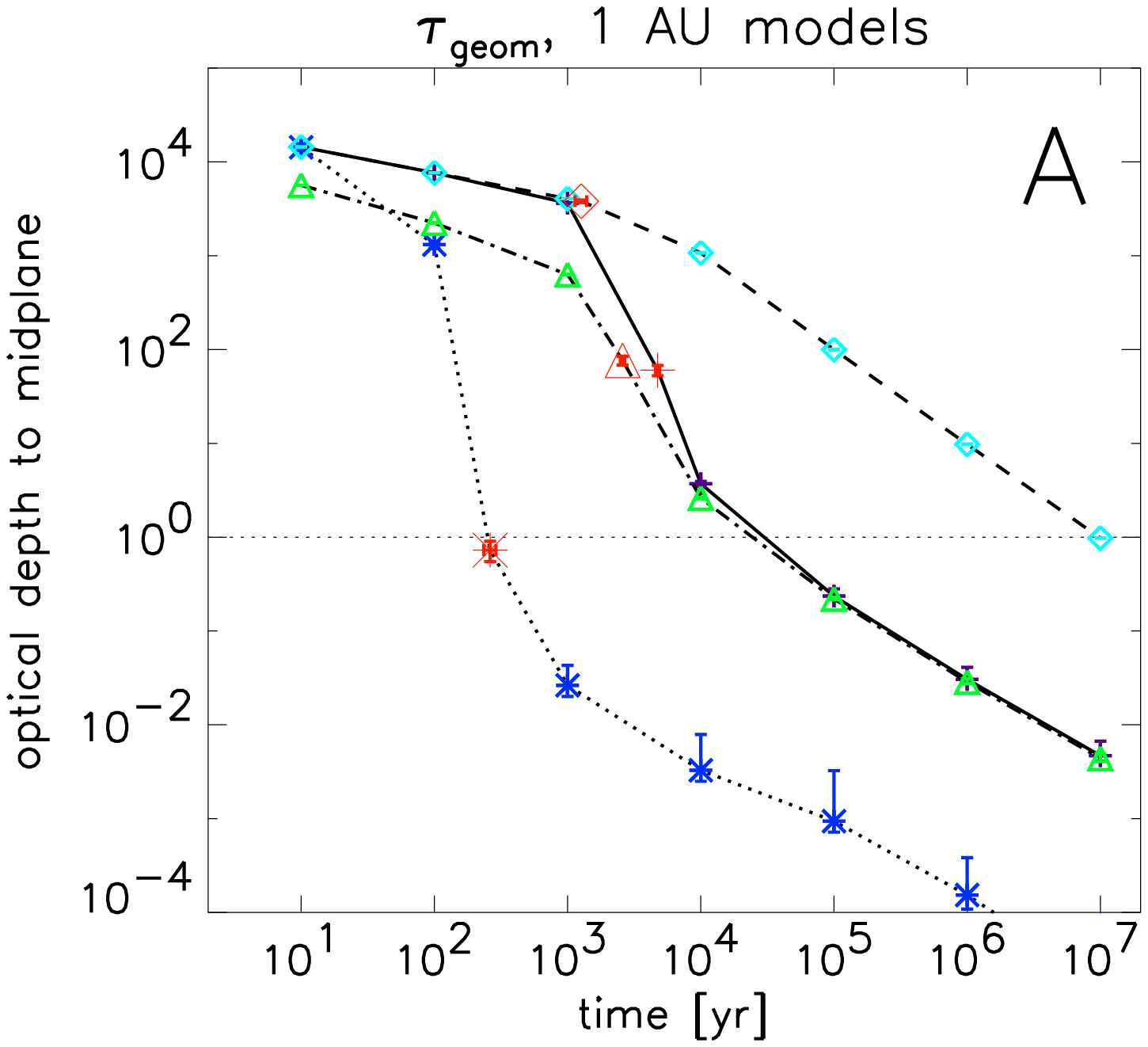}\includegraphics[clip=true,scale=0.5]{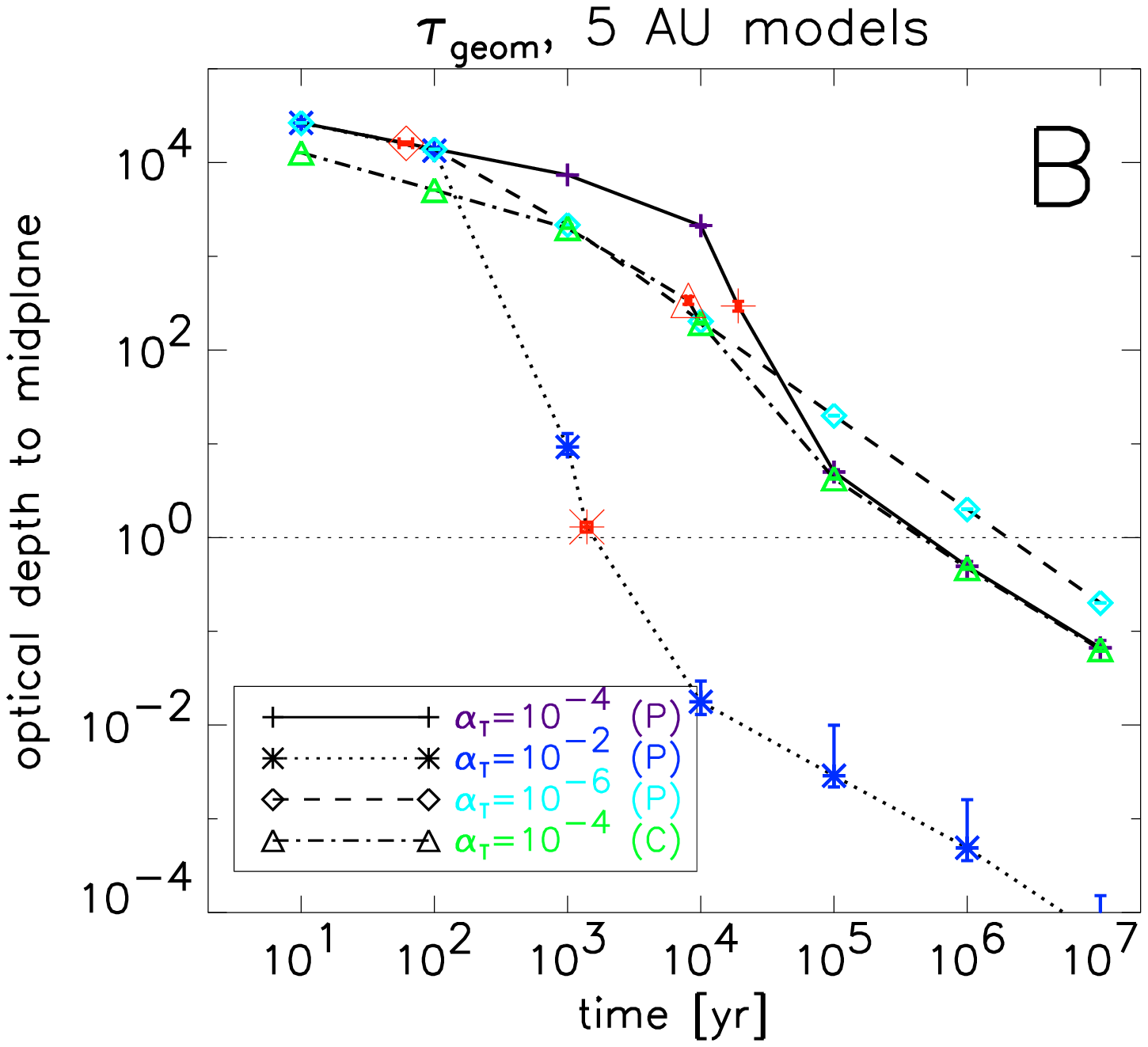}\\
  }
  \caption{Geometrical optical depth to the mid-plane as function of time for the 1 AU models (left) and the 5 AU models (right). The optical depth is computed at every logarithmic interval in time and also at the point where particles rain-out (red symbol). For illustrative purposes the points are connected by lines. Error bars denote the spread throughout the 50 runs of each model.}
  \label{fig:opacgeom}
\end{figure*}
It now becomes clear that the internal structure of particles, represented here by the \changed{enlargement} parameter $\psi$, is a variable of key importance in models of dust-aggregation. To illustrate this point further, Fig.\ \ref{fig:growthmodels} compares the `evolution curve' of our (default $\alpha_\mathrm{T} = 10^{-4}$, 1 AU) model with those of compact, PCA and CCA aggregation (grey curves). The superimposed black curve connects the $(m,\psi)$ values of the most massive particle resulting from the collision model. The small, erratic structure in this curve corresponds to the fact that the particles fluctuate stochastically during the simulation. Furthermore, since we compare single particles here, instead of a (weighted) mean of the distribution as, e.g., in Fig.\ \ref{fig:evolcurves}, a fixed point in the figure corresponds to one particular friction time and lines of equal friction times lie parallel to the dashed lines indicating $\tau=\tau_\mathrm{rain}$ and $\tau=t_\mathrm{s}$. For the specific choice of $\alpha_\textrm{T}=10^{-4}$ we see that $\tau_\mathrm{rain}$ comes prior to $t_\mathrm{s}$ and velocities therefore remain modest. Still, in our collision model the coagulation will reach a point, indicated by a cross in Fig.\ \ref{fig:growthmodels}, at which some collisions, having the right mass ratio and relative velocity, are energetic enough to cause compaction ($E > E_\mathrm{roll}$), which effectively halts any further increase in porosity. However, due to the earlier extensive build-up of porosity in the fractal regime, the particle distribution now evolves to larger sizes as compared to the compact models (Fig.\ \ref{fig:massfunc1}), causing the rain-out masses to be orders of magnitudes higher (Table\ \ref{tab:rainout}). Thus, a major part of the growth takes place in the nebula phase. Large, porous particles are quickly produced, stay in the nebula mixed with the gas and only settle when they are sufficiently compacted, e.g., by energetic collisions (this paper) or shocks (see below).

The curve of our model, with its characteristic bending point due to compaction, is a direct result of treating porosity as a dynamic variable that is altered by the collisional process. In all other models in Fig.\ \ref{fig:growthmodels}, on the other hand, compaction is not incorporated, resulting in straight lines when the growth is in the fractal regime. In the PCA/CCA fractal models compaction is of course \textit{a priori} ruled-out. However, the pre-assumed absence of compaction in the PCA/CCA models is consistent with the low collision energies in these limiting cases. In CCA, friction times barely grow ($\tau_\textrm{f} \propto m^{0.05}$), the $t_\textrm{s}$ `threshold' is not reached and relative velocities vanish for similar particles. In PCA, on the other hand, relative velocities are higher, but the collision energy is now suppressed by the reduced mass. Thus, if the coagulation process would (for some reason) be restricted to these limiting growth modes, aggregates will not restructure. Fig.\ \ref{fig:growthmodels} shows how this affects the overall coagulation process. It shows, e.g., that compact grains rain-out at $\sim 10^{-4}\ \mathrm{g}$, `PCA grains' at $\sim 10^{-2}\ \mathrm{g}$, porosity-evolving particles of our model at $\sim 1\ \mathrm{g}$, and `CCA particles' will grow forever! It is clear that the porosity evolution of collisional agglomerates is of decisive influence on the coagulation process. Modelling the porosity evolution in combination with a microphysical collision model is therefore a key requirement for a full understanding of the first stages of planet formation.

To quantify the effects of coagulation on the appearance of the disk, we have calculated optical depths to the mid-plane in the MSN. As a first order approximation, the vertical structure is to be taken of constant density and extends over one scaleheight. We assume that this layer is represented by the particles of our simulation and that the rain-out particles (which we do not follow) are below it, i.e., at the mid-plane regions of the disk. The geometrical optical depth (i.e., at visible/UV-wavelengths) \textit{to} the mid-plane is then calculated as
\begin{equation}
  \tau_\mathrm{geom} = H_\mathrm{g} \int \mathrm{d}m\ f(m) \pi a^2.
\end{equation}
Results are given in Fig.\ \ref{fig:opacgeom} for the 1 AU and 5 AU models. In Fig.\ \ref{fig:opacgeom}A it is seen that the $\alpha_\mathrm{T}=10^{-6}$ models stay optically thick for most of the disk's evolution. Note also that the $\alpha_\mathrm{T}=10^{-4}$ porous models (solid lines) and the $\alpha_\mathrm{T}=10^{-4}$ compact models (dashed-dotted line) do not deviate much in $\tau_\mathrm{geom}$. This shows again the dual effects of porosity on the population: it increases the geometrical cross-section, yet it also speeds up the coagulation, causing more mass to be `locked' inside large particles. At 5 AU (Fig.\ \ref{fig:opacgeom}B), the timescales are longer and the disks only becomes optically thin after $\sim 10^6\ \mathrm{yr}$ when $\alpha_\mathrm{T} \lesssim 10^{-4}$. In the $\alpha_\mathrm{T} = 10^{-2}$ models, evolution to optical thinness is very fast at both radii. It is clear that, within the frame-work of these models, the inner regions of protoplanetary disks are rapidly depleted of small grains unless $\alpha_\textrm{T} \sim 10^{-6}$ or less.

There is a further serious issue hidden here. All our models show a rapid decline of the (sub-)micron size population on a timescale of $\sim 10^3\ \mathrm{yr}$. This is a well-known problem in models for grain coagulation in protoplanetary environments: the densities are high enough for coagulation to proceed rapidly \citep{2005A&A...434..971D}. Furthermore, the turbulence induced relative velocities promote collisions between particles with different friction times, i.e., between small and larger particles. In contrast, observations reveal the presence of copious amounts of small grains in the disk-photospheres of isolated Herbig AeBe stars and T-Tauri stars, pointing toward the presence of small grains on timescales of some $10^6\ \mathrm{yr}$ \citep{2004Natur.432..479V,2005A&A...437..189V,2003A&A...409L..25M,2006astro.ph..2041N}. This discrepancy implies a continuous replenishment of (sub)micron-sized grains. In particular, it may reflect the importance of vaporisation and condensation processes continuously forming fresh, small grains. Likely, this vaporisation and condensation would be localised in the hot and dense, inner regions of the disk and these grains would then have to be transported upwards and outwards to the disk photosphere through diffusion processes. The high degree of crystallinity of silicates in the inner few AU of protoplanetary disks also points toward the importance of condensation processes in these environments \citep{2004Natur.432..479V}. Likewise, the presence of crystalline silicates in the cold outer regions of protoplanetary disks has been attributed to large scale mixing of materials in these environments \citep{2002A&A...384.1107B,2004A&A...413..571G}. Alternatively, the replenishment of small grains is through collisional fragmentation. These energetic collisions could take place either in the gaseous nebula due to high relative velocities driven by a high $\alpha_\mathrm{T}$, or in the mid-plane regions with the subsequent upwards diffusion of small grains. While it might be difficult to sustain $\alpha_\textrm{T} \gtrsim 10^{-2}$ over a prolonged period of time, fragmentation in the mid-plane regions seems viable since more massive particles will reside here. Furthermore, if the mid-plane becomes dust-dominated ($\rho_\mathrm{d} > \rho_\mathrm{g}$), shear-turbulence will develop, further augmenting the collisional energies \citep{1993Icar..106..102C}. 

Further constraints on the collisional growth of grains in protoplanetary disks are provided by the solar system record; specifically, the chondrules and Ca-Al-rich Inclusions (CAI), which dominate the composition of primitive meteorites. These millimetre-sized igneous spherules are high-temperature components that formed during transient heating events in the early solar system. We realise that the cm-sized fluff-balls formed in our porous coagulation models are in the right mass-range of these meteorite components. It is tempting to identify these fluff-balls as the precursors to the chondrules and CAIs. We might then envision a model where the collisional evolution is terminated by the flash-heating event, for example a shock or lightning, which leads to melting and the formation of a chondrule and subsequent immediate settling. During the settling phase the chondrule may acquire a dust rim by sweeping up small dust grains or other unprocessed fluff-balls still suspended in the nebula \citep{2004Icar..168..484C}. One key point to recognise here is that chondrules show a spread in age of a few million years \citep{2005ASPC..341..953W}, which indicates that the collisional grain growth process takes place over a much longer timescale than our models would predict (see above).

In this study we have focused on the agglomeration driven by random motions -- either Brownian or turbulent -- high up in the nebula. Growth is then presumed to be terminated once the aggregate has compacted enough to settle. At that point, the aggregate will \changed{drop-down} about one scaleheight after which further growth must occur for further settling to \changed{continue}. In reality, instead of the simple two-component picture of nebula and mid-plane presented in this work, the nebula will acquire a stratified appearance \citep{1995Icar..114..237D}, where larger particles with higher friction times have smaller scaleheights. Collisional evolution models should be able to include this stratified nature of the disk. \changed{This stratification also extends in the radial direction due to radial drift motions and turbulent diffusion.} Incorporation of these motions into the Monte Carlo code will be challenging. We expect that the study presented here can serve as the basis for incorporating realistic grain growth in hydrodynamical models, likely in the form of well-selected `collision-recipes'. An obvious step would be to include collisions that exceed the $E_\mathrm{frag}$ limit in the collision model. Note, for example, that in the \citet{1997ApJ...480..647D} terminology what we have called `fragmentation' should in fact be sub-divided in a continuous range of aggregate disruptions. At first monomers will be lost and only if $E \gg E_\mathrm{frag}$ are aggregates completely shattered. Other extensions to the model are to allow for a distribution of monomer sizes and to use monomers of different chemical composition. Both will affect the critical energy for restructuring, $E_\textrm{roll}$. However, with an increasing number of parameters characterising the collision, it is worthwhile to verify experimentally -- either through laboratory experiments or through detailed numerical calculations -- which are of prime importance. 

\section{\label{sec:concl}Conclusions}
We have presented a model that incorporates the internal structure of aggregates as a variable in coagulation models. We used the \changed{enlargement} parameter $\psi$ to represent the internal structure. It is seen that the internal structure is key to the collisional evolution, since it crucially affects the dust-gas coupling. However, in the model presented in Sect.\ \ref{sec:model} $\psi$ is not a static variable. It is altered by the collisional process and we have constructed simple recipes to include this aspect in coagulation models.

Next, we have applied the new collision model to the collisional evolution of the turbulent protoplanetary disk, until particles rain-out to the mid-plane. Our main conclusions are:
\begin{itemize}
  \item With the treatment of porosity as a variable, three different regimes can be distinguished: fractal growth, compaction and fragmentation \changed{\citep{2004ASPC..309..369B}} . These regimes are also reflected in the collisional evolution of the size distribution: fractal growth (mostly Brownian motion), compaction (growth accelerates) and rain-out. 
  \item The collisional evolution of our porosity-evolving model is quantitatively different from PCA/CCA aggregation models in which porosity can be parameterised by a fixed exponent. Therefore, a microphysical collision model is a key requirement for coagulation models.
  \item Due to the porous evolution, particles upto dm-sizes can be suspended in the gaseous nebula, orders of magnitude larger in mass than in models of compact coagulation. Therefore, chondrule precursors could have had their origin in regions above the mid-plane.
  \item If $\alpha_\mathrm{T} < 10^{-2}$, no fragmentation occurs in the gaseous nebula. Therefore, if $10^{-6} < \alpha_\textrm{T} < 10^{-2}$, the inner nebula will become optically thin on timescales of $\sim 10^4\ \mathrm{yr}$, unless an influx of small grains takes place.
\end{itemize}

\begin{acknowledgements}
  CWO likes to show his appreciation to Carsten Dominik and Dominik Paszun for extensive discussion on the manuscript. These greatly helped to improve the quality of the paper, especially regarding Sect.\ \ref{sec:model}. Carsten Dominik is also thanked for suggesting Figs.\ \ref{fig:fractalfig} and \ref{fig:evoltime} and Dominik Paszun for useful assistance with creating Fig.\ \ref{fig:fractalfig}. Finally, the referee, J\"urgen Blum, is thanked for providing a thorough and helpful review.
\end{acknowledgements}

\bibliographystyle{aa}
\bibliography{5949publ}
\appendix
\section{List of symbols}
\begin{table}[h!]
  \caption{List of symbols.}
  \centering
  \begin{tabular}{lp{5cm}}
  \hline \hline
  $\Delta \mathrm{v}$     &   relative velocity\\
  $\Sigma$                &   surface density\\
  $\Omega$                &   angular (Keplerian) rotation frequency\\
  $A$                     &   \changed{geometrical} cross-section (1 particle)\\
  $C$                     &   collision rate \\
  $E$                     &   collision energy \\
  $E_\textrm{restr}/E_\textrm{max-c}/E_\textrm{frag}$  & \changed{energy limits for particle restructuring/ maximum-compression/ fragmentation} \\
  $H_\mathrm{g}/h$        &   gas/particle scale height \\
  $K$                     &   collision kernel \\
  $N$                     &   number of monomers (Sect.\ \ref{sec:model}); number of particles (Sect.\ \ref{sec:mccode}) \\
  $N_\mathrm{c}$          &   number of contacts\\
  \changed{$N_\textrm{s}$} &   \changed{number of distinct particles (number of species)}\\
  $R$                     &   heliocentric radius\\
  $Re$                    &   Reynolds number\\
  $T$                     &   temperature\\
  $V/{\cal V}$            &   particle/simulated volume \\
  $\alpha_\mathrm{T}$     &   turbulent viscosity parameter \\
  $\xi$                   &   critical displacement \\
  $\gamma$                &   specific surface energy \\
  $\delta$                &   area-mass exponent ($A\sim m^\delta$) in fractal regime  \\
  $\kappa$                &   number of coagulations\\
  $\mu$                   &   reduced mass\\
  $\nu_\mathrm{m}/\nu_\mathrm{T}$ &   molecular/turbulent viscosity\\
  $\psi$                  &   \changed{enlargement parameter} \\
  $\rho_\mathrm{d}/\rho_\mathrm{g}/\rho_\mathrm{s}$  & dust/gas/specific density \\ 
  $\sigma$                &   \changed{collision} cross-section \\
  $\tau_\mathrm{f}$       &   friction time \\
  $\tau_\mathrm{rain}$    &   \changed{friction time at which particles are removed from simulation due to rain-out}\\
  $a/a_0$                 &   radius (size)/monomer size \\
  $c_\mathrm{g}$          &   thermal sound speed\\
  $g_i$                   &   number population of species $i$.\\
  $k_\mathrm{B}$          &   Boltzmann constant \\
  $\ell$                  &   eddy length scale\\
  $m$                     &   mass \\
  $f$                     &   particle distribution function (number density spectrum) \\
  $r$                     &   random deviate \\
  $t$                     &   time \\
  $t_s/t_0$               &   \changed{turn-over timescale of smallest/largest eddy} \\
  $\mathrm{v}$            &   velocity \\
  $\mathrm{v_\textrm{s}}/\mathrm{v_0}$ & \changed{velocity smallest/largest eddies} \\
  \hline
  \end{tabular}
\end{table}
\end{document}